\documentclass[conference]{IEEEtran}
\IEEEoverridecommandlockouts

\usepackage{fancyhdr}
\fancypagestyle{firstpage}{
    \fancyhf{}
    \fancyfoot[C]{\footnotesize
    \copyright\ 2026 IEEE. Personal use of this material is permitted. Permission from IEEE must be obtained for all other uses, in any current or future media, including reprinting/republishing this material for advertising or promotional purposes, creating new collective works, for resale or redistribution to servers or lists, or reuse of any copyrighted component of this work in other works.}

}
\usepackage{cite}
\usepackage{amsmath,amssymb,amsfonts}
\usepackage{graphicx}
\usepackage{textcomp}
\usepackage{xcolor}
\def\BibTeX{{\rm B\kern-.05em{\sc i\kern-.025em b}\kern-.08em
    T\kern-.1667em\lower.7ex\hbox{E}\kern-.125emX}}

\usepackage{color, colortbl}
\usepackage{amsmath}

\usepackage{graphicx}
\usepackage{url}
\usepackage{tcolorbox}
\usepackage{booktabs}
\usepackage{algorithm}
\usepackage[noend]{algpseudocode}
\usepackage{subcaption}
\usepackage[normalem]{ulem}
\usepackage{multirow}
\usepackage{listings}
\usepackage{xcolor}
\usepackage{xspace}
\usepackage{enumitem}
\usepackage{pifont}
\usepackage{hyperref}
\usepackage{xurl}
\hypersetup{
    colorlinks = true,
    linkcolor = blue,
    citecolor = blue
}
\tcbuselibrary{breakable}
\newcommand{\Tx}{\mathcal{T}}
\newcommand{\EVMCall}{\mathcal{C}}

\definecolor{verylightgray}{rgb}{0.97,0.97,0.97}
\definecolor{darkgreen}{rgb}{0.0, 0.5, 0.0}

\lstdefinelanguage{Solidity}{
  keywords={contract, function, uint, uint256, calldata, address, require, assert, revert, returns, if, else, public, bool, payable, external},
  keywordstyle=\color{blue}\bfseries,
  sensitive=true,
  comment=[l]{//},
  morecomment=[s]{/*}{*/},
  morestring=[b]",
}

\def\toolname{CuEVM\xspace}

\def\ityfuzz{ItyFuzz\xspace}
 
\def\fuzztoolname{CuEVM-fuzz\xspace}

\algnewcommand{\LeftComment}[1]{\State \(\triangleright\) #1}

\newcommand{\Description}[1]{}

\begin{document}

\title{CuEVM: Scaling Smart Contract Execution on GPUs}
\author{
\IEEEauthorblockN{
Nhut-Minh Ho\textsuperscript{1},
Tien Tuan Anh Dinh\textsuperscript{2},
Quang Trung Ta\textsuperscript{1},
Fredrik Svantes\textsuperscript{3},
Xiaokui Xiao\textsuperscript{1},
Beng Chin Ooi\textsuperscript{4}
}
\IEEEauthorblockA{
\textsuperscript{1}\textit{National University of Singapore}, Singapore \quad
\textsuperscript{2}\textit{Deakin University}, Australia \\
\textsuperscript{3}\textit{Ethereum Foundation}, Switzerland  \quad \textsuperscript{4}\textit{Zhejiang University}, China  \\
minhhn@comp.nus.edu.sg, anh.dinh@deakin.edu.au, taqt@nus.edu.sg,\\ fredrik@ethereum.org, xkxiao@nus.edu.sg, ooibc@zju.edu.cn
}
}

\maketitle
\thispagestyle{firstpage}

\begin{abstract}
Smart contracts on Ethereum blockchain process transactions worth billions of US dollars daily. They run on top of Ethereum Virtual Machine (EVM) which executes transactions sequentially. This execution model creates a performance bottleneck, especially for new blockchain workloads with large numbers of transactions, such  as security testing. We present CuEVM, a GPU-native and semantically complete EVM execution engine that treats transaction execution as a massively parallel data-processing problem. It employs novel data structures and GPU-aware scheduling to accelerate execution. Unlike prior works that transpile the contract and rely on a host CPU for many operations, CuEVM supports complete EVM semantics natively on the GPU. It passes 100\% of the Ethereum VMTests and 96.2\% of the full execution-test suite. On a single commodity GPU, CuEVM executes over 8 million ERC-20 transfers per second, outperforming existing baselines by up to two orders of magnitude. We build a smart contract fuzzer, called \fuzztoolname{}, on top of \toolname{}. The fuzzer achieves over a million end-to-end executions per second, up to $23\times$ and $4.8\times$ higher than the state-of-the-art CPU-based and GPU-based baselines, respectively. 
\end{abstract}

\section{Introduction}
\label{sec:intro}

Smart contracts enable complex blockchain applications that process high-value transactions. For example, decentralized finance (DeFi) applications on Ethereum have over 150 billion USD in total locked value~\cite{defillama}. The Ethereum Virtual Machine (EVM) executes transactions sequentially, even when the transactions are independent, creating a throughput bottleneck~\cite{wood2014ethereum}. This bottleneck is critical for new blockchain applications that process a large number of transactions, and most of the transactions are independent and can be executed in parallel. Examples of such applications include transaction simulation~\cite{chen2026mevisor, meng2026houston} and coverage-guided fuzzing~\cite{MAUchen2024towards,ItyfuzzShou2023}.

Our goal is to design an EVM engine that executes parallel transactions over thousands of GPU threads.
There are two challenges in realizing our goal. The first is supporting complete EVM semantics under a massively parallel execution model. A recent work on GPU-based EVM execution, namely MAU~\cite{MAUchen2024towards}, transpiles smart contract bytecode to PTX assembly to emulate execution. But it supports only a subset of the EVM semantics~\cite{wood2014ethereum}, omitting critical features such as external calls, precompiled contracts, contract creation, and rollbacks. It also restricts the world state to a single contract without cross-contract interactions. MAU fails more than 80\% of the official Ethereum tests and must fall back to a host CPU reference EVM, which reintroduces the sequential execution. The second challenge is reducing memory access latency and CPU-GPU communication overhead. In particular, memory access is sensitive to data layout~\cite{shanbhag2020study}. Thus, EVM states and auxiliary data structures must be carefully designed to avoid overhead. MAU, for example, uses separate data buffers to track fuzzing code coverage for different GPU threads, therefore incurring high overhead. 

We present \toolname{}, a semantically complete EVM engine for conflict-free workloads that runs natively on GPUs. \toolname{} addresses the first challenge by executing EVM bytecode following the yellow paper~\cite{wood2014ethereum}. It is compatible with existing EVM implementations, and passes 100\% of the Ethereum VMTests and 96.2\% of the full execution-test suite. \toolname{} addresses the second challenge by adopting a columnar layout for the EVM states, batching tens of thousands of transactions per kernel, and using a GPU-aware transaction scheduler.

Using \toolname{} for our primary use case, we build \fuzztoolname{}, the first GPU-native fuzzer that migrates the entire fuzzing loop to GPUs. Our evaluation shows that \fuzztoolname{} can perform up to 1.1 million end-to-end fuzzing executions per second on an Nvidia RTX 5000 Ada. It detects existing CVEs $300\times$ faster than a state-of-the-art fuzzer (Smartian). On 500 large smart contracts collected from Etherscan~\cite{etherscan}, it identifies $1.75\times$ more integer overflow bugs and $2\times$ more leaking ether bugs than Smartian, and $1.67\times$ more reentrancy bugs than ItyFuzz. In real-world testing on 20,000 popular contracts on Ethereum and Binance Smart Chain, \fuzztoolname{} detects 115 projects with high-risk bugs.
In summary, we make the following contributions:

\begin{itemize}[nosep,noitemsep,leftmargin=1.5em]

    \item We introduce \toolname, the first GPU-native EVM that is %
    fully adhering to the EVM specification\cite{wood2014ethereum}. It passes all the core functionality tests in the official Ethereum test suite. The code is available at \url{https://github.com/sbip-sg/CuEVM}.

    \item We present a novel design that incorporates columnar-style memory layout and GPU-aware transaction scheduling that minimizes GPU-related overheads. Our microbenchmarks show that \toolname{} can execute up to 8.3 million independent ERC-20 transfers per second.

    \item We introduce \fuzztoolname{} that leverages {\toolname} for smart contract fuzzing. \fuzztoolname{} performs up to 1.1 million end-to-end fuzzing executions per second. It detects new high-risk bugs in 115 recent real-world contracts.%

    \item We discuss another use case of {\toolname}, namely attack imitation~\cite{blockchain-immitation-game}, in which \toolname{} enables efficient search for rescue transactions upon detection of on-chain security incidents.

\end{itemize}

The next section presents relevant background.
Section~\ref{sec:time-critical-ronin-use-case} describes the motivating example, followed by the overview of \toolname{} in Section~\ref{sec:system_overview}.
Section~\ref{sec:CuEVM} provides the design and implementation of \toolname.  Section~\ref{sec:fuzzing_implementation} discusses the design of {\fuzztoolname{}}.
Section~\ref{sec:experiment} presents the experimental results. Section~\ref{sec:related_work} discusses the related works before Section~\ref{sec:conclusion} concludes.

\section{Background}
\label{sec:background}

\subsection{Blockchain and Smart Contract}

A blockchain is a replicated, append-only, and tamper-evident ledger maintained by a set of mutually distrustful nodes that agree on its contents through a consensus protocol~\cite{mohan2017tutorial,dinh2018untangling}. From a data-management standpoint,  blockchains share notable similarities with, but also key differences from, distributed databases~\cite{ruan2021blockchains}, a dichotomy that has motivated hybrid blockchain-database systems~\cite{ge2022hybrid}. A common abstraction decomposes a blockchain into four layers: application, execution engine, data, and consensus for performance benchmarking~\cite{dinh2017blockbench}. \toolname{} speeds the execution layer up by harnessing the parallel processing power of GPUs.

Applications are enabled through smart contracts, and Ethereum is the most popular blockchain supporting smart contracts. All nodes in Ethereum agree on a world state $\sigma$ --- the states of all accounts in the blockchain. The world state is a mapping between account address $a$ and its state. For each 160-bit address $a$, the account state $\sigma[a]$ consists of four components:
\begin{equation}
   \sigma[a] = \langle \mathtt{nonce}, \mathtt{balance}, \mathtt{code}, \mathtt{storage} \rangle
   \label{eq:state_account}
\end{equation}
where  $\mathtt{code}$ and $\mathtt{storage}$ are for smart contract accounts.

A user interacts with Ethereum by submitting a transaction $\mathcal{T}$, which updates the current world state $\sigma$ to $\sigma'$ via $\sigma' = \Upsilon(\sigma, \mathcal{T})$, where $\Upsilon$ is the state transition function that executes $\mathcal{T}$. Each transaction $\mathcal{T}$ has various fields and is signed by the sender with five key fields relevant to transaction execution as below:
\begin{equation}
   \mathcal{T} = \langle \mathtt{sender}, \mathtt{receiver}, \mathtt{data}, \mathtt{value}, \mathtt{gasLimit}, \ldots \rangle
\end{equation}

When $\mathcal{T}$'s $\textsf{receiver}$ is a smart contract address, the EVM passes $\mathcal{T}$'s \textsf{data}, which contains the encoded high-level function and arguments following Application Binary Interface (ABI) encoding~\cite{zheng2020application}. The \textsf{value} encodes the transaction value in Ethereum while the \textsf{gasLimit} determines the upper bound on execution complexity. The EVM executes the contract's bytecode to perform state transition logic in $\Upsilon$ following EVM semantics~\cite{wood2014ethereum}.
The execution may involve multiple accounts and multiple smart contract calls. Once it completes successfully, the state changes are materialized to a new world state $\sigma'$.

Smart contracts are written in higher-level languages such as Solidity before being compiled to bytecode and deployed on the blockchain. They are susceptible to bugs and deployment errors that may lead to catastrophic financial loss. %

\subsection{Ethereum Virtual Machine}
\label{sec:background_EVM}
Ethereum Virtual Machine (EVM) is the engine that executes smart contracts on Ethereum and on other EVM-compatible blockchains, such as Polygon and Binance Smart Chain. The EVM is a stack-based virtual machine with a word size of 256 bits. It executes instructions represented as opcodes, embedded in the bytecode and indexed by a program counter ($pc$). Each transaction $\Tx$ instantiates an EVM instance to apply the state transition function $\Upsilon$ to the world state $\sigma$. The $\text{EVM}$ first launches an initial
contract call to execute the target contract bytecode. This initial call may spawn a number of sub-calls, depending on the contract logic and the runtime state.
Each call maintains a separate \emph{call context} ($\EVMCall$) that captures the runtime
state of the EVM. We represent $\EVMCall$ as:
\begin{equation}
\EVMCall = \langle \underbrace{env, bytecode, data, value}_{\text{static}} \mid
\underbrace{stack, memory, \sigma, pc, gas}_{\text{dynamic}} \rangle
\label{eq:call_context}
\end{equation}
where the components are defined as follows:
\begin{itemize}[nosep, noitemsep, leftmargin=1.5em]
\item \textbf{Static state} (immutable during $\EVMCall$ execution):
      environment $env$, smart contract $bytecode$, call $data$, call $value$
\item \textbf{Dynamic state} (mutable during $\EVMCall$ execution):
     $stack$, $memory$, world state $\sigma$,
      program counter $pc$, remaining $gas$
\end{itemize}

The environment $env$ of each call includes the call depth, the sender, receiver, the state address, call gas limit, transaction origin, and blockchain-related states such as block number, block time, etc. The sender and receiver of each sub-call may differ from the transaction’s originator (\textsf{tx.origin}) and the original target contract. The state address is required for special opcodes such as \textsf{delegatecall} or \textsf{callcode} where the receiver of the call differs from the contract owning the code or the storage.

This execution model maps directly onto database transaction processing: the shared world state $\sigma$ is the persistent data, and the contract logic together with the EVM internals defines the state transition that each transaction applies. The EVM executes transactions sequentially, each mutating $\sigma$ to a new version, and a group of transactions forms a block that is finalized by the consensus layer.

\subsection{GPU-Accelerated Smart Contract Execution}

\subsubsection{GPUs}
\label{sec:background_GPU}

Parallel computing has evolved from multi-threaded to massively parallel models on special hardware like GPUs~\cite{dally2021evolution}.
A GPU program is called a kernel (written in CUDA for Nvidia GPUs~\cite{cook2012cuda}), which can run thousands of lightweight threads in parallel. The threads are grouped into blocks, each containing warps of 32 threads. These warps are the instruction execution units, which are scheduled to the streaming multiprocessors on the GPU. The whole warp can only execute one instruction at a time. When threads in the warp wait for each other's diverged instructions, thread divergence incurs additional overhead. Each thread has its own registers for local variables, while all threads within a block share a small but faster shared memory. All threads in the kernel can access a slower global memory, which is gigabytes in size. Thread memory access patterns and execution path are important factors that affect GPU program performance. %

\begin{figure*}
    \centering
    \includegraphics[width=0.7\linewidth]{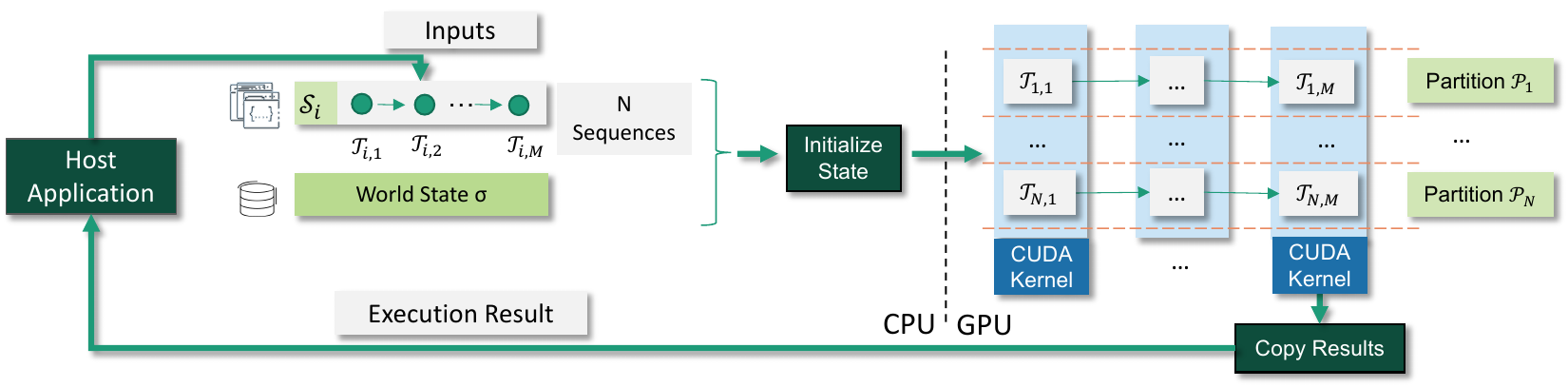}
    \caption{Overview of \toolname{}}
    \Description{}
    \label{fig:evm-instance-overview}
\end{figure*}
\subsubsection{Smart contract execution on GPUs}

To the best of our knowledge, there is no existing general-purpose EVM execution engine on GPUs. For a specialized engine serving a specific purpose of fuzzing smart contracts, MAU~\cite{MAUchen2024towards} introduces a GPU-accelerated fuzzer with a smart contract emulator running on GPUs. It transpiles EVM bytecode to PTX assembly to simulate execution on GPUs and utilizes Ityfuzz~\cite{ItyfuzzShou2023} on the CPU side to complete each fuzzing loop.
From the execution-engine perspective, MAU suffers from two drawbacks: incomplete
semantics and low GPU utilization. We postulate that a semantically complete EVM is necessary not only for fuzzing use cases as in MAU but also for general parallel workload where correctness is as important as speed.

\section{Motivation}
\label{sec:time-critical-ronin-use-case}
We observe that there are emerging blockchain applications that process a large number of transactions, and most of the transactions can be executed in parallel. One example is parallel clients that re-execute blocks to validate optimistic schedules~\cite{gelashvili2023block, lin2025parallelevm}. Another example involves  executing synthetic transactions to find vulnerabilities~\cite{MAUchen2024towards}, to build anomaly detection models~\cite{meng2026houston}, or to find rescue transactions~\cite{blockchain-immitation-game}.
These applications are currently limited by the EVM because the latter executes transactions sequentially.

\subsection{Smart Contract Fuzzing}
Smart contracts process transactions worth billions of dollars. Thus, implementation bugs or security vulnerabilities can lead to catastrophic financial loss~\cite{tophack}. Fuzzing is a widely adopted technique to find bugs and vulnerabilities in smart contracts, inspired by the success of fuzzers in traditional
software testing~\cite{AFL} and database systems~\cite{gao2025fucci}. A (greybox) fuzzer executes a large number of transactions. It reads feedback from each execution to decide which transactions are
\textsf{interesting} and to generate new transactions~\cite{nguyen2020sfuzz,Smartianchoi2021,torres2021confuzzius}. Listing~\ref{lst:sample_contract} shows an example from the Ronin Bridge contract, where the bug is an unauthorized withdrawal that becomes reachable after calling \texttt{initializeV4()}. To trigger this bug, the fuzzer must drive the contract execution into this state and then issue \texttt{submitWithdrawal($\cdot$)} with the corresponding call data. Because the space of callable functions, their arguments, and their orderings is large, the fuzzer mutates transactions to make progress under the guidance of a feedback metric, for example, code coverage. In summary, fuzzing requires executing a large number of candidate transactions, most of which are independent from each other, and the throughput of the engine determines how quickly bugs are found.

 \begin{lstlisting}[caption={Simplified contract from the Ronin Bridge incident.}, label={lst:sample_contract}]
contract MainchainGatewayV3 {
  // function initializeV3()
  // function initializeV4()

  function submitWithdrawal(Transfer.Receipt calldata _receipt, Signature[] calldata _signatures) external virtual whenNotPaused () {
    // other logic
    uint256 quantity = _receipt.info.quantity;
    address tokenAddr = _receipt.mainchain.tokenAddr;
    if(_reachedWithdrawalLimit(tokenAddr,quantity)){
      revert();} // withdrawal limit check.

    // flawed verification with _totalOperatorWeight = 0
    // execute withdrawal quantity from tokenAddr
  }
  // ... other functions
}

 \end{lstlisting}

\subsection{Other Use Cases}
In August 2024, the Ronin Bridge was exploited for over 12 million USD\footnote{The fund was rescued by whitehats with a bounty}. The root cause was a recently upgraded contract with misconfigured state variables, allowing unauthorized withdrawals (Listing~\ref{lst:sample_contract}). The incident unfolded as follows:
\begin{itemize}[nosep,noitemsep,leftmargin=1.5em]
 \item Block 20468438: the contract was misconfigured. One block later (approximately 12 seconds), an exploit transaction with the value of 34 USD was recorded.
 \item Block 20468679: an optimized transaction with the value of 10 million USD (3,996 ETH) was recorded approximately 49 minutes later.
 \item Block 20468866: the bridge was paused to prevent further loss 38 minutes after the first major loss.
\end{itemize}

The incident reveals that the attacker went through multiple exploits to extract the most value from the vulnerability. This gives a window for security researchers to respond to the incident. In particular, the researchers solve the following problem: \textit{given a vulnerability, find the exploit that extracts the most value}. Such an exploit is called {\em rescue transaction}, as it can front-run the real exploits from the attacker to rescue the asset~\cite{blockchain-immitation-game,your-exploit-is-mine, whitehatstats}.
In the Ronin bridge example, the parameter \texttt{quantity} at bytes 324–356 of the transaction data and the daily limit of 4,000 ETH determine the maximum extractable amount. Without knowledge of the logic, finding the rescue transaction requires random mutation of transaction data and measuring the extracted value. \toolname{} helps find transaction parameters in one second that could have extracted 9,000 USD more than the recorded transaction.

\toolname{} is useful beyond fuzzing and incident response. In particular, it enables MEV search\cite{chen2026mevisor}, speed up block validation by parallelizing execution of independent transactions within the block\cite{gelashvili2023block}.

\section{Overview}
\label{sec:system_overview}

\toolname{} is a semantically complete EVM engine that executes batches  of transactions entirely in GPUs. Figure~\ref{fig:evm-instance-overview}  shows how it interacts with a {\em host application} running on CPU. The host application maintains persistent data, schedules transaction batches, and implements the application logic.

\subsection{Host Application}
A host application runs multiple rounds of batched execution until it reaches a target output or state. In each round, it schedules a batch of transaction sequences with a world state, then processes the execution results (or feedback) to drive the next round. This workflow applies to multiple use cases. 
\begin{itemize}[nosep,noitemsep,leftmargin=1.5em]
\item \textbf{Smart contract fuzzing}: the host application collects execution feedback in terms of code coverage, and uses it to generate the next batch of transactions.
\item \textbf{Incident response}: the host application refines the search regions of the transaction parameters after each round.
\item \textbf{MEV search:} the host application tracks the best transaction ordering with the most extractable value, and generates new ordering after each round. 
\end{itemize}
Adapting a CPU-based application to \toolname{} requires a GPU-aware scheduler and input generator that keep the GPU fully utilized. We discuss these components for smart contract fuzzing in Section~\ref{sec:fuzzing_implementation}.

\subsection{Execution Model}

The host application runs in iterations. Each iteration generates a batch of $N$ transaction sequences, where each sequence contains up to $M$ transactions:
\begin{equation}
\text{Batch}=\{\mathcal{S}_i\}_{i=1}^N,\qquad
\mathcal{S}_i=(\mathcal{T}_{i,1},\mathcal{T}_{i,2},\dots,\mathcal{T}_{i,M}),
\label{eq:batch_defintion}
\end{equation}
where \(\mathcal{T}_{i,j}\) denotes transaction \(j\) of the sequence \(i\)

\toolname{} adopts the shared-nothing, partitioned execution model: each sequence runs on its own partition $\mathcal{P}_i$ that holds an isolated version $\sigma_i$ of the world state, analogous to GPU-based transaction execution over independent partitions~\cite{he2011high}. This is the ideal, fully partitioned case. Some use cases, however, require data shared across partitions. For clarity we first present the general-purpose engine with $N$ partitions and read-only shared state, which covers the majority of use cases; we defer the handling of mutable shared state to the fuzzing use case in Section~\ref{sec:fuzzing_implementation}.

\toolname{} executes the batch by a sequence of $M$ CUDA kernels. The $j$-th kernel applies the $j$-th transaction of every sequence in parallel, that is, $\sigma_i \leftarrow \Upsilon(\sigma_i, \mathcal{T}_{i,j})$ for $i \in [1, N]$, following the Ethereum Yellow Paper~\cite{wood2014ethereum}. Each partition $\mathcal{P}_i$ is initialized with its own version $\sigma_i$ of the world state before the first kernel and persists this version throughout the batch execution. At the end of each iteration, the results are transferred back to the host for persistent storage before the next iteration begins.

\toolname{} is built for high parallelism, supporting hundreds of thousands of transactions in parallel, bounded mainly by device memory. Because a warp executes one instruction at a time across its 32 threads, the dominant performance risk is control-flow divergence within a warp. \toolname{} therefore exposes a scheduling interface to the host application: the host should group transactions with similar control flow into the same warp. The engine is agnostic to how this grouping is produced and stays generic for possible applications on the host; each application supplies a GPU-aware adaptor that emits warp-friendly batches. We describe the concrete adaptor for fuzzing, which assigns one contract function per warp and varies only its inputs, in Section~\ref{sec:fuzzing_implementation}. This is finer-grained than MAU, which forces an entire kernel of thousands of threads to execute a single function and is therefore far more restrictive for the host.

\subsection{CPU-GPU Communication}
\label{sec:cpu_gpu_communication}
A naive design transfers a large amount of data between the host and device at every transaction, which is dominated by data-movement cost over the host-device interconnect~\cite{pearson2019evaluating_gpu_mem}. \toolname{} instead exchanges data only at batch boundaries, once per direction per iteration: host-to-device when a batch of $N$ sequences is dispatched, and device-to-host when its results are collected. The result payload is application-defined. In fuzzing it is a list of bugs or coverage-increasing sequences, while in MEV search or incident response it is the sequence with the highest yield. This batched scheme follows from the observation that most transaction-execution use cases need only the final output of a sequence rather than its intermediate states. Further reductions require application-specific optimization and compression, for which we present techniques for the fuzzing use case in Section~\ref{sec:fuzzing_implementation}.

\section{Design and Implementation}
\label{sec:CuEVM}

This section describes the core component \toolname{} in detail, which implements EVM semantics on GPU, executes N sequences of transactions in parallel on N partitions, each containing a version of the world state.

\subsection{Data Structures and Data Layout}

\toolname{} runs thousands of transactions in parallel, each on its own EVM instance. Their combined execution state behaves like a large collection of records that are read and updated concurrently. We adopt two ideas from data systems to make this efficient. A columnar layout (Section~\ref{sec:coalesced_memory_intro}) stores the same field of all instances contiguously~\cite{abadi2008column}. An optimistic undo-logging scheme (Section~\ref{sec:world_state_rollback}) handles transaction rollback without snapshotting or versioning the world state at every call. Each instance executes against its own version of the world state, so instances do not interfere and can run in parallel without coordination.

\subsubsection{Memory access pattern and thread divergence}
\label{sec:coalesced_memory_intro}
CUDA programs are sensitive to memory access patterns, where adjacent threads accessing adjacent memory locations can maximize effective memory bandwidth~\cite{fauzia2015characterizing_gpu_coalesced}. This scheme is called \textit{coalesced memory access}~\cite{gpu_coalesced_memory}. Because memory requests within a warp are grouped before issuance, scattered requests introduce extra memory transactions, increasing delay and reducing throughput. Similarly, thread divergence within a warp (Section~\ref{sec:background_GPU}) incurs extra instructions and stalls as threads wait for divergent paths to complete.

For coalesced memory access, we arrange the logical data elements of adjacent EVM instances contiguously in memory. When threads access the same logical data structure (e.g., the first item on each EVM's stack), they access adjacent memory locations.
For example, at the beginning of execution, neighboring threads execute the same \textsf{PUSH} instruction and update the top of their respective stacks. This benefits from coalesced access, as the first entries of the stacks are allocated contiguously in memory.

To reduce thread divergence, we expose the transaction-scheduling interface to the host program. The ideal case is when all 32 threads in a warp execute the same instruction at each step. For example, in the fuzzing use case, the host can schedule transactions that target the same contract function, so threads in a warp follow similar execution paths and divergence is reduced.

\label{sec:memory_allocation}
\begin{figure}
    \centering
    \includegraphics[width=0.9\linewidth]{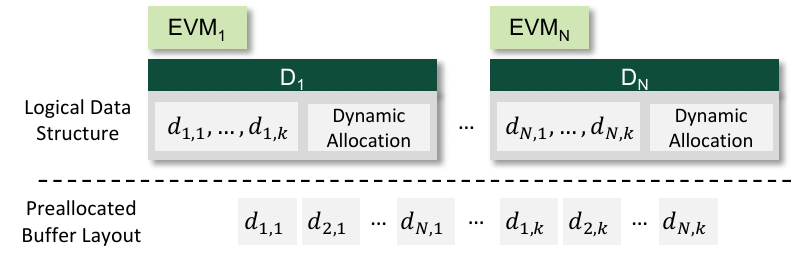}
    \caption{Columnar data layout for coalesced memory access}
    \Description{}
    \label{fig:data_structure}
\end{figure}
\subsubsection{Call context}
Each EVM instance creates a call context $\EVMCall$ (Section~\ref{sec:background_EVM}) composed of several data structures. We distinguish two kinds. Fixed-size components are elementary values of known width, such as the 256-bit integers for the gas tracker and the transaction value, and the 64-bit values for the program counter and the call-data size. Dynamic-size components carry a data segment whose length grows and shrinks during execution, namely the \texttt{memory}, the \texttt{stack}, and the call \texttt{data}. Each dynamic-size component also has fixed-size fields, such as its current size and the stack head, which we treat like the other fixed-size components.

We lay out all fixed-size components in the columnar form. For example, the \texttt{gas} value of instance $\text{EVM}_i$ is stored next to the \texttt{gas} value of instance $\text{EVM}_{i+1}$, and a warp that reads the \texttt{gas} of consecutive instances issues coalesced memory requests. A fixed-size component is the degenerate case of one element per instance, whereas a dynamic-size component has multiple elements per instance and changes during execution. We represent a dynamic-size component with the abstraction $D = [d_1, \ldots, d_k]$ and lay out its elements with the same columnar scheme. Because the world state is more complex, each instance stores a pointer to its own version of the world state $\sigma$ that it executes against. The world-state design supports the rollback mechanism in Section~\ref{sec:world_state_rollback}.

\subsubsection{Columnar data layout}
\label{sec:evm_instance_data_structure}
We use one general layout for the runtime state of all instances, designed for coalesced access. Given $N$ instances, each instance $\text{EVM}_i$ holds its own version $D_i$ of a logical data structure, where each element has the type appropriate to its component, such as a 256-bit integer for stack contents. We preallocate $k$ elements for each $D_i$, so $D_i = [d_{i,1}, \ldots, d_{i,k}]$, giving a buffer of size $k \times N$ laid out as in Figure~\ref{fig:data_structure}. Rather than store each instance's structure contiguously, we store the corresponding element of all instances contiguously: first every instance's element $d_{\cdot,1}$, namely $d_{1,1}, d_{2,1}, \ldots, d_{N,1}$, then every element $d_{\cdot,2}$, and so on through the $k$-th elements $d_{1,k}, d_{2,k}, \ldots, d_{N,k}$. Equivalently, viewing the buffer as an $N \times k$ matrix with one row per instance and one column per element position, we store it in column-major order. This is the struct-of-arrays, or columnar, organization used in column-oriented data systems~\cite{abadi2008column}, and it is what makes a warp's access to one element position across consecutive instances coalesced. If a structure outgrows its $k$ elements, we fall back to dynamic allocation at a higher cost~\cite{winter2021dynamic_allocate}. We use this layout for all major call-context structures except the EVM memory, whose pages we store contiguously per instance for efficiency. The value of $k$ is configurable and defaults to a size that serves common ERC-20 transfers without dynamic allocation.

\subsection{Sub-Calls}
When a call context $\EVMCall_{j}$ makes a sub-call, it creates a child context $\EVMCall_{j+1}$ with its \textsf{depth} incremented by one relative to $\EVMCall_{j}$, up to the maximum depth of 1024. When $\EVMCall_{j+1}$ terminates, the EVM applies the corresponding termination logic and restores the parent context $\EVMCall_{j}$.

Each sub-call creates its own runtime data structures, as defined in Equation~\ref{eq:call_context}, while preserving the parent calls' data for later restoration. This adds a depth dimension to the columnar layout: the $N \times k$ layout conceptually becomes an $N \times \textsf{depth} \times k$ tensor, since every instance can hold a stack of live call contexts. Materializing it directly would scale memory with the call-depth bound of 1024, at $k$ elements per depth. We instead flatten the depth dimension into the per-instance buffer, with each depth at a distinct contiguous offset range, as we illustrate next with the EVM's stack of a call $\EVMCall_{j}$:

\begin{equation}
    stack = \langle \textsf{base\_ptr}, \textsf{offset}, \textsf{dyn\_ptr}, \textsf{size}\rangle
\end{equation}
where \textsf{base\_ptr} points to the beginning of the preallocated data segment, \textsf{offset} denotes where this call's stack content begins, \textsf{dyn\_ptr} points to the dynamically allocated segment if one exists, and \textsf{size} is the current stack size. When creating a new context $\EVMCall_{j+1}$ from $\EVMCall_{j}$, we compute the child offset $\textsf{offset}_{j+1}$ from the parent's $\textsf{offset}_{j}$ and $\textsf{size}_{j}$ as follows:
\begin{equation}
\textsf{offset}_{j+1}
= \min\!\left( \textsf{offset}_{j} + \textsf{size}_{j},\; \textsf{buffer\_size} \right)
\end{equation}
where \textsf{buffer\_size} is the per-instance buffer size preallocated in Figure~\ref{fig:data_structure}, that is, the value $k$. This lets sub-calls share one buffer at distinct starting offsets while preserving the parent's data within the same buffer. On return from a sub-call, the child's portion of the buffer is cleared to ensure correctness. This design, combined with the columnar layout, enables coalesced memory accesses across all call depths and relevant data structures. Adjacent threads with similar workloads are likely to execute the same instructions and access the same logical data at the same call depth, where the corresponding data is stored contiguously in memory. We use this design for all relevant call-context data structures.

\subsection{World State and Rollback Mechanism}
\label{sec:world_state_rollback}
\toolname{} maintains a single world state shared among all sub-calls of the same EVM instance. Except for the contract bytecode, the account data fields in Equation~\ref{eq:state_account} are allocated following the layout in Section~\ref{sec:evm_instance_data_structure}. We preallocate a buffer of a configurable size ($x$ accounts $\times$ $y$ storage slots) and implement the storage as two arrays, one of keys and one of values, using the layout in Figure~\ref{fig:data_structure}. Each value $V$ is a tuple $\langle v_{curr}, v_{org}, v_{warm}\rangle$, where $v_{curr}$ is the current 256-bit value, $v_{org}$ is the value before the transaction, and $v_{warm}$ indicates the warm state of the slot. These fields are required to compute gas usage correctly~\cite{EIP2929}. We allocate memory dynamically when the storage exceeds the preallocated buffer or when new accounts are created during execution.

When a contract call fails due to executing a \textsf{revert} opcode or encountering runtime exceptions, the world state is rolled back to the state before the call.
State updates are committed to storage only if the call terminates gracefully, that is, on \textsf{stop} or \textsf{return}. When \textsf{revert} occurs at the sub-call level, the transaction is still committed, and only the changes made by the reverted sub-calls are discarded when they return. Because most blockchain transactions are expected to complete successfully, we adopt an optimistic rollback strategy: all writes are applied directly to a single world state and recorded in an undo log. On a revert, \toolname{} replays the log in reverse to restore the original values for the current call. This avoids the cost of snapshotting or versioning the world state at every call and pays the rollback cost only on the rare revert.

\subsection{Other EVM Semantics}

\textbf{Opcode semantics and execution model.}
\toolname{} supports all EVM opcodes, each implemented as a separate function of the EVM instance following the Yellow Paper~\cite{wood2014ethereum}. It includes a gas meter that tracks the execution cost of each opcode, including the dynamic gas components that can only be resolved at runtime, such as the costs of message data and memory expansion.

\textbf{256-bit integer arithmetic.}
Since CUDA does not natively support 256-bit integers, we implement a 256-bit arithmetic library from scratch. It is adapted from the Cooperative Groups Big Numbers (CGBN) library~\cite{cgbn} to support thread-level parallelism and finite-field arithmetic.

\textbf{Precompiled contracts and hashing functions.}
Precompiled contracts are special addresses exposing commonly used functions that contracts can call. They are analogous to User-Defined Functions in databases. \toolname{} implements all precompiled contracts of the EVM Shanghai version, adapting existing libraries~\cite{sha,bigintC,pyecc} and the reference implementation in Go Ethereum~\cite{go-ethereum}.

\section{\fuzztoolname{}: Coverage-Guided Fuzzing}
\label{sec:fuzzing_implementation}
We design a coverage-guided fuzzer, called \fuzztoolname{}, on top of CuEVM. Fuzzing is selected because it is widely used for securing smart contract, therefore \fuzztoolname{} demonstrates \toolname{}'s practicality. In addition, fuzzing is execution bound, which serves as an interesting workload to  demonstrate \toolname{}'s performance and scalability.   
\subsection{Smart Contract Fuzzing}
\label{sec:fuzzing_smartcontract}

Coverage-guided fuzzing is an execution-bound workload that stresses every part of the engine, which makes it a natural first use case for \toolname{}. Execution dominates the cost, because a campaign tests many transactions before a bug is revealed, so parallel execution is the established way to scale fuzzing~\cite{medusa,MAUchen2024towards}. The workload, therefore, maps directly onto the batched, data-parallel model of \toolname{}. In each iteration, the fuzzer forms a batch of $N$ sequences, executes them in parallel on private clones of the world state, and adds the interesting sequences to a shared corpus. We build \fuzztoolname{} on this model. A lightweight controller on the CPU handles persistent storage and scheduling, and the GPU runs execution together with three resident modules, namely a mutator for mutating transaction data, a coverage tracker for tracking progress, and a set of bug oracles for detecting bugs.

\subsection{Batched Execution over Private State}
In \fuzztoolname{}, each EVM instance executes a sequence on a private clone of the world state. Within a batch there are no cross-transaction conflicts, so the engine needs no concurrency-control protocol for the transaction state, and the work reduces to layout and scheduling. The world state, the coverage map, and the bugs list are persistent on the GPU and only the world state is reinitialized from a snapshot at the start of each batch, which keeps the working set on the device and avoids per-iteration transfer.

Algorithm~\ref{alg:gpu_batched_execution} shows the full loop. The CPU controller drives three steps per iteration. \textsc{NextBatch} draws a batch of $N$ sequences from the corpus (Step \ding{172}). \textsc{ExecuteBatch} runs that batch on the GPU and returns the interesting events $\mathcal{O}$ (Step \ding{173}). The controller then decodes each event back into a corpus update through \textsc{Reconstruct} (Step \ding{174}), which rebuilds the originating sequence on the host rather than copying it back from the device. The loop typically repeats until the time budget is spent or a bug is found. %

\textsc{ExecuteBatch} runs entirely on the GPU and is where the private-state model takes shape. It first resets the $N$ private clones $\sigma_i$ from the resident snapshot, which re-initializes the world state for the new batch. It then advances the batch through $M$ kernel launches, where launch $j$ moves every sequence forward by one position. Inside a launch the $N$ threads run in parallel, and thread $i$ does three things for its sequence. It mutates the next transaction with \textsc{Mutate}, applies the EVM transition function $\Upsilon$ to its own clone $\sigma_i$, and reports any feedback through \textsc{TrackEvent}. Because $\sigma_i$ stays resident across the $M$ launches, a sequence of length $M$ runs to completion before returning to the host. When all threads finish, \textsc{AggregateResults} collects the per-thread findings into one buffer for the controller. Both \textsc{TrackEvents} and \textsc{AggregateResults} operate on the shared coverage and bug state and are detailed in Section~\ref{sec:coverage_tracking_impl}.

\begin{figure}[t]
    \centering
    \includegraphics[width=0.92\linewidth]{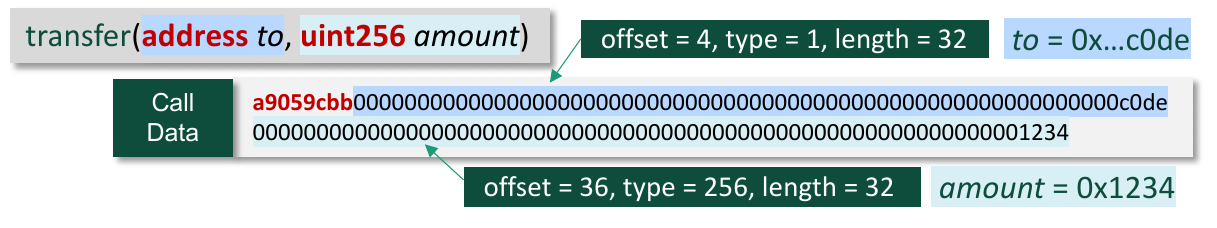}
    \caption{Sample configuration $\gamma$ for ERC-20 transfer}
    \Description{}
    \label{fig:datamarker_example}
\end{figure}

Before execution, each transaction is mutated by a type-aware mutator that respects the ABI argument types embedded in the contract bytecode, which avoids transaction early reverts when those checks fail. A fuzzable transaction carries a configuration $\Gamma = \langle seed, [\gamma_1, \ldots, \gamma_n] \rangle$, where each $\gamma_k = \langle \textit{offset}, \textit{type}, \textit{length} \rangle$ marks a mutable argument by its byte offset, type, and length in the transaction data (Figure~\ref{fig:datamarker_example}). The \textsc{Mutate} procedure seeds a pseudo-random generator from $\Gamma.seed$ and applies standard AFL value mutators to each marked argument according to its type, such as bit and byte flips, arithmetic mutation, and dictionary insertion~\cite{AFL}.
 \begin{algorithm}[t]
\caption{End-to-end GPU-accelerated fuzzing loop}
\label{alg:gpu_batched_execution}
\small
\begin{algorithmic}[1]
\State $Corpus \gets$ initial corpus \Comment{CPU controller}
\State $\sigma \gets$ initial world state snapshot
\While{within time budget}
    \State $\text{Batch} \gets \Call{NextBatch}{Corpus}$ \Comment{Step \ding{172}, CPU}
    \State $\mathcal{O} \gets \Call{ExecuteBatch}{\text{Batch}, \sigma}$ \Comment{Step \ding{173}, GPU}
    \For{$o$ in $\mathcal{O}$} \Comment{Step \ding{174}, CPU}
        \State $Corpus \gets Corpus \cup \{\Call{Reconstruct}{o}\}$
    \EndFor
\EndWhile
\vspace{0.3em}
\Procedure{ExecuteBatch}{Batch, $\sigma$} \Comment{Runs on GPU}
    \For{$i$ in $1..N$} \Comment{Reset private clones per batch}
        \State $\sigma_i \gets \sigma$ \Comment{Snapshot isolation}
    \EndFor
    \For{$j$ in $1..M$} \Comment{$M$ kernels, one per sequence position}
        \For{$i$ in $1..N$} \Comment{$N$ GPU threads, parallel}
            \State $\mathcal{T}_{i,j} \gets \Call{Mutate}{\mathcal{T}_{i,j}, \Gamma_{i,j}}$
            \State $\sigma_i \gets \Upsilon(\sigma_i, \mathcal{T}_{i,j})$ \Comment{EVM semantics}
            \State \Call{TrackEvent}{$\sigma_i$} \Comment{Algorithm~\ref{alg:track_events}}
        \EndFor
    \EndFor
    \State \Return \Call{AggregateResults}{ } \Comment{Algorithm~\ref{alg:track_events}}
\EndProcedure
\vspace{0.3em}
\Procedure{Mutate}{$\mathcal{T}, \Gamma$}
    \State $\textit{rand} \gets \Call{SeedRandom}{\Gamma.seed}$
    \For{$\gamma$ in $\Gamma$}
        \State $\mathcal{T} \gets \Call{ArgMutate}{\mathcal{T}, \gamma, \textit{rand}}$ \Comment{Type-aware}
    \EndFor
    \State \Return $\mathcal{T}$
\EndProcedure
\end{algorithmic}
\end{algorithm}

\subsection{Concurrent Feedback over Shared State}
\label{sec:coverage_tracking_impl}
\subsubsection{Shared fuzzing state}
Most of the engine state is private, but two structures are shared across all instances and updated concurrently by thousands of GPU threads, namely the coverage map and a set of oracle-specific buffers. We treat every update to these structures as a concurrent update to shared aggregate state and resolve it with CUDA atomic operations rather than locks~\cite{gpu_atomic_function}.

The coverage map records execution feedback in the form of branch coverage and storage access. It is an array of $2^{16}$ words, following the design common to existing smart contract fuzzers~\cite{nguyen2020sfuzz,ItyfuzzShou2023,MAUchen2024towards}. For a branch event, each word packs two signals, an 8-bit branch distance in its high bits and a 24-bit hit count in its low bits,
\begin{equation}
    \textsc{Cov}[\mathit{eid}] = \underbrace{\mathit{distance}}_{\text{bits}[31{:}24]} \ \Vert\ \underbrace{\mathit{hitcount}}_{\text{bits}[23{:}0]}.
    \label{eq:mapentry_encoding}
\end{equation}
The hit count records how often an event has been observed across the campaign, so a word that is still zero marks a branch or storage access reached for the first time. The branch distance measures how close a conditional was to flipping. At the comparison that precedes a conditional jump, we record the gap between the compared operands, and a smaller gap means the input is closer to taking the branch it currently misses. This matters because EVM conditions often compare 256-bit values, where a random input either takes a branch or misses it with no signal in between, and the distance turns that all-or-nothing outcome into a progress the fuzzer can improve over time~\cite{ItyfuzzShou2023,nguyen2020sfuzz}. We use the bit count of the distance instead of its absolute 256-bit value.

The event identifier $\mathit{eid}$ selects the entry in the coverage map. For a branch it combines the current program counter $pc$, the taken destination $pc_{dst}$, and the account address $a$,
\begin{equation}
\mathit{eid} = (pc \ll 1 \oplus pc_{dst}) \oplus a,
\label{eq:event_id}
\end{equation}
and for a storage access it combines $pc$, the slot key, and $a$. The collision property of this encoding scheme has been analyzed in~\cite{wang2019branchkeyAFL}. The address term $a$ is to track coverage across different smart contracts in the world state.

The oracle-specific buffers hold the extra evidence to confirm and report a specific finding, such as minimized historical call data that can confirm whether the attacker can make an arbitrary call. Oracle writes into this buffer by incrementing an atomic index, so that concurrent findings do not collide.

\subsubsection{Reporting feedback to shared state}
Algorithm~\ref{alg:track_events} shows how each thread reports feedback. \textsc{TrackEvent} records a new branch or storage event by atomically incrementing the word and testing whether the prior hit count was zero, so the thread that first reaches an event is the one that reports it. \textsc{TrackDistance} records progress toward a missed branch by an atomic maximum on the distance field. We encode the value as $255 - \mathit{distance}$, so a shorter distance produces a larger stored value, and \textsc{AtomicMax} keeps an input only when it improves on the best distance seen so far. We use \textsc{AtomicMax} because the distance occupies the most significant bits, so the comparison is decided by the distance field, and the lower hit count bits cannot change its outcome. Both atomics return the old value, allowing the current thread to determine whether it actually updated the value. For extra data such as the full event $o$ that must be copied back to the CPU, each thread writes its findings to a private buffer $\mathcal{R}$ during execution. Once all threads finish, \textsc{AggregateResults} merges these buffers into the shared output buffer $\mathcal{O}$ using an atomic index. When multiple threads aggregate into the same buffer, \textsc{AtomicAdd} increments the index $count$ exactly once per thread, returning a unique index into the shared buffer for updating. %

\begin{algorithm}[t]
\caption{Concurrent feedback over shared state on GPU}
\label{alg:track_events}
\small
\begin{algorithmic}[1]
\State $\mathcal{R} \gets \emptyset$ \Comment{Private buffer per thread}
\Procedure{TrackEvent}{$o$} \Comment{Hit branch or storage event}
    \State $t \gets \Call{AtomicAdd}{\textsc{Cov}[o.\mathit{eid}], 1}$
    \If{$t\ \&\ \texttt{0xffffff} == 0$} $\mathcal{R} \gets \mathcal{R} \cup \{o\}$ \Comment{New event}
    \EndIf
\EndProcedure
\vspace{0.3em}
\Procedure{TrackDistance}{$o, \mathit{distance}$} \Comment{Missed branch}
    \State $\mathit{cmp} \gets (255 - \mathit{distance}) \ll 24$
    \State $t \gets \Call{AtomicMax}{\textsc{Cov}[o.\mathit{eid}], \mathit{cmp}}$
    \If{$\mathit{cmp} > t$} $\mathcal{R} \gets \mathcal{R} \cup \{o\}$ \Comment{Distance reduced}
    \EndIf
\EndProcedure
\vspace{0.3em}
\State $\mathcal{O} \gets [o_1, \ldots, o_m]$ ; $\mathit{count} \gets 0$ \Comment{Shared buffer and index}
\Procedure{AggregateResults}{} \Comment{Parallel over threads}
    \For{$o \in \mathcal{R}$}
        \State $\mathit{idx} \gets \Call{AtomicAdd}{\mathit{count}, 1}$
        \If{$\mathit{idx} < m$} $\mathcal{O}[\mathit{idx}] \gets o$
        \EndIf
    \EndFor
\EndProcedure
\end{algorithmic}
\end{algorithm}

\subsection{Minimizing Host-device Data Movement}
\label{sec:cpu-gpu-communication}
We minimize host-device traffic in both directions for performance. The mutation configurations and the initial world state snapshot are cached on the GPU to avoid frequent transfers. We schedule each transaction sequence to a warp of 32 threads that share one configuration, and transfer a single index $idx$ per warp to select the cached configuration together with a per-batch base seed, where thread $k$ derives its seed as $seed + k$. On the return path we avoid copying mutated transaction data back by implementing an equivalent pair of pseudo-random generators on both the CPU and GPU. The controller reconstructs an interesting sequence by replaying the same mutation from the interesting thread indices ($i,j$) in a batch. Each reported event is therefore compact,
\begin{equation}
\mathcal{O} = [o_{i,j}], \quad o_{i,j} = \langle \mathtt{seq}, \mathtt{eid}, \mathtt{data} \rangle,
\label{eq:event_data_structure}
\end{equation}
where $\mathtt{seq}$ is a placeholder the controller fills by replay, $\mathtt{eid}$ identifies the bug or coverage entry, and $\mathtt{data}$ carries any event-specific payload.

\subsection{Bug Oracles and Controller}
\label{sec:bug_oracle_impl}
\fuzztoolname{} implements its bug oracles natively on the GPU. We support integer overflow and underflow~\cite{Smartianchoi2021}, unauthorized self-destruct, leaking ether, and arbitrary external call~\cite{solidity_security_concern}, and reentrancy that violates the check-effect-interaction pattern~\cite{solidity_security_concern}. We select these because they cover high-risk classes that existing tools also detect, appear in common benchmarks~\cite{Smartianchoi2021,ferreira2020smartbugs}, and are verifiable from source code or from the exploit sequence.

The controller builds on Medusa~\cite{medusa}, a property-based fuzzer from Trail of Bits, which lets us reuse its configuration format, corpus, coverage reporting, and sequence generators with less engineering effort. We extend its generator with function dependency information from Slither~\cite{feist2019slither}. For a function $f$, we derive $F_{\textit{dep}}(f)$, the functions that write variables $f$ reads, and $F_{\textit{infl}}(f)$, the functions that read variables $f$ writes. We prepend a sample from $F_{\textit{dep}}(f)$ and append a sample from $F_{\textit{infl}}(f)$ so that generated sequences exercise state dependencies. The controller schedules transactions so that each group of $k$ GPU threads targets the same smart contract function. By default, $k$ equals the warp size of 32. The controller can also load an arbitrary initial state, which supports stateful and cross-contract campaigns. The address term $a$ in the event identifier of Equation~\ref{eq:event_id} already separates coverage and bugs per deployed contract to support cross-contract tracking natively.

\section{Evaluation}
\label{sec:experiment}
\subsection{Methodology}

We first evaluate the performance and correctness of \toolname{} in executing transactions in the Ethereum Execution Tests format \cite{eth-tests}. We then evaluate the performance of \fuzztoolname{} and incident response application. Our experimental artifact is available online~\cite{cuevm-artifact}.

\textbf{Metrics.} We use transactions per second (TPS) to measure transaction execution throughput. We measure fuzzing effectiveness in terms of code coverage and the number of bugs found over time. In particular, we use unique instructions covered for code coverage~\cite{MAUchen2024towards, ItyfuzzShou2023, Smartianchoi2021}.
In EVM, unique instructions are identified by their program counter (\texttt{pc}) values.
For a fair comparison, we run each baseline under its best-performing configuration and adopt the common \texttt{pc} counting method, documenting any changes.

\textbf{Workloads and Datasets.}
We evaluate \toolname{} along two axes. The workloads (ERC-20 transfers and Uniswap swaps) characterize raw execution throughput under popular transaction types and different complexities. They represent the backbone of DeFi applications and account for the majority of Ethereum traffic. The datasets (LabeledDataset, EtherScan500, RecentContract, and PopularContract) are collections of real contracts used to evaluate \fuzztoolname{}. \textsf{LabeledDataset} comprising two bug types: Integer Overflow/Underflow Bug (IB, 58 contracts) and Reentrancy (RE, 19 contracts) from Smartian~\cite{Smartianchoi2021}.
The IB contracts are reported CVEs, and the RE contracts are from the Smartbugs dataset~\cite{ferreira2020smartbugs}.
We use three unlabeled datasets to evaluate different aspects of \fuzztoolname{}. First, for direct comparison with existing works, we use \textsf{EtherScan500} dataset from Smartian, which consists of the 500 largest and most popular smart contracts collected from Etherscan. This dataset evaluates performance on complex real-world contracts that all baselines can process successfully. We further construct \textsf{RecentContract} dataset, which contains 3,691 contracts recently deployed on Ethereum, predominantly compiled with Solidity version 0.8. Finally, to demonstrate practical security relevance, we conduct a large-scale scan on the \textsf{PopularContract} dataset, consisting of 20,000 popular contracts with active user activity over the past two years on Ethereum and Binance Smart Chain. %

\textbf{Experiment setup.} We run all experiments on a workstation with an Intel® Core™ i9 processor 14900 running Ubuntu 24.04 LTS. The machine is equipped with 128GB RAM and an NVIDIA RTX 5000 Ada with 32GB memory. %
Except the large-scale smart contracts scanning experiment, we ran each experiment 3 times and report the average.

\subsection{Correctness, Throughput, and Scalability}
\subsubsection{Execution correctness}
 We integrated CuEVM with GoEVMLab \cite{goevmlab} to compare our EIP-3155 output with Geth on Ethereum's Test Suite~\cite{eth-tests} to evaluate its correctness. The test suite consists of over 14,000 tests divided into smaller sets to test different aspects of EVM semantics. A test is passed if \toolname{}'s trace matches Geth on the gas usage and the full stack content in each executed instruction. The results show that CuEVM passes 100\% of the VMTests subset, which contains the core EVM semantics, and it passed 96.2\% of all the tests. Most failures are due to minor gas differences in corner cases, which typically involve reverts near the call depth limit and trace printing timed out due to the extraction of large stack content for all instructions.
These failures come from stress-test-style cases and CUDA's printing limitation, thus they do not affect practical applications such as fuzzing, where transactions stay below the limit and printing is disabled. %

We instrumented MAU for correctness comparison. Because heavy instrumentation disrupts MAU operation, we relaxed the check: %
a test passed when MAU matches Geth in the opcode and the first stack item on each executed instruction. We disregard gas comparisons, as MAU does not support recent gas calculations such as EIP-2929. Although this estimates an upper bound for correctness, MAU only passes 18.7\% of the tests. Most failures stem from external calls and contract creation, which are common in real-world smart contracts. %

\subsubsection{Bytecode execution throughput}
\begin{figure}[]
    \centering
    \includegraphics[width=0.8\linewidth]{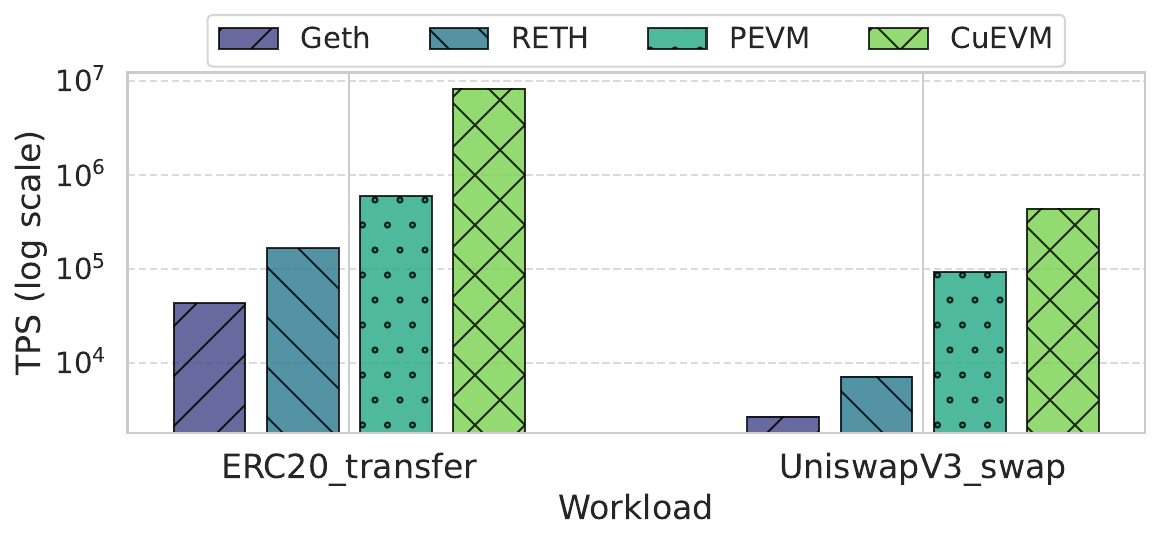}
    \caption{Comparison of execution throughput}
    \Description{}
    \label{fig:uninstrumented_evm_bench}
\end{figure}

We conduct microbenchmarks to measure the peak throughput of \toolname{} under non-conflicting transaction workloads. We compare against three EVM implementations: GoEthereum (Geth)~\cite{go-ethereum}, Rust-based EVM (RETH)~\cite{reth}, and RETH-based Parallel EVM (PEVM)~\cite{pevm}. All measurements only record the bytecode execution logic, without the memory initialization and transfer time, which account for less than 0.1 second in \toolname. 
We validate \toolname against Geth by comparing EIP-3155 traces on this workload and confirm it matches the traces produced by Geth. We defer MAU comparison to the fuzzing use case because it lacks support for external calls. For CuEVM, we scale the workload up to 512,000 transactions and report the best result. Figure~\ref{fig:uninstrumented_evm_bench} shows the results. For ERC20 transfers, \toolname achieves 8.3 million TPS. For Uniswap transactions involving many external calls, \toolname{} achieves the throughput of around 441,000 TPS.

\subsubsection{Scalability}
\begin{figure}[]
    \centering
    \includegraphics[width=0.8\linewidth]{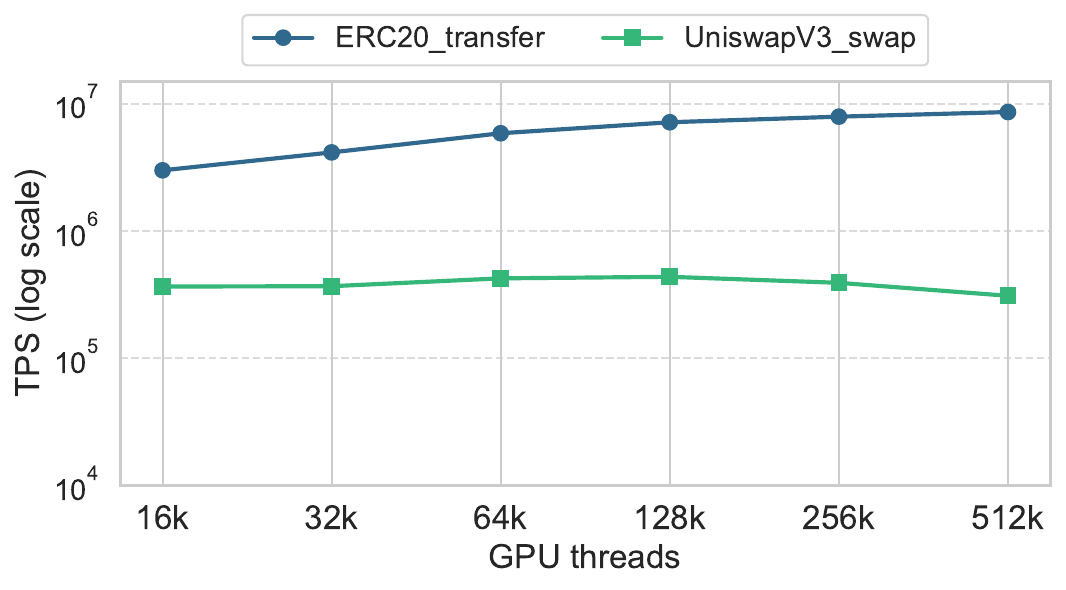}
    \caption{\toolname{} throughput scaling}
    \Description{}
    \label{fig:scalability_cuevm}
\end{figure}
Figure~\ref{fig:scalability_cuevm} reports the scalability of \toolname{} as the number
of GPU threads grows from 16k to 512k, with one transaction per thread. The two
workloads scale differently, reflecting the data-layout choices in
Section~\ref{sec:CuEVM}. ERC20 fits the default preallocated buffer, therefore the throughput increases to over 8M TPS, ultimately bounded by memory capacity: at 512k threads it consumes about 24.9 GB. Uniswap peaks at
128k threads and then slightly declines. Its many external
calls and deeper call stack use the dynamic allocated memory region which is the source of contention as the number of threads
grows. We note that the buffer size is configurable, thus it is possible to profile the target workload and size
the buffer accordingly to reduce this contention. 
\subsection{\fuzztoolname{}}
We compare \fuzztoolname{} against four other state-of-the-art smart contract fuzzers: Smartian~\cite{Smartianchoi2021}, ItyFuzz~\cite{ItyfuzzShou2023}, MAU~\cite{MAUchen2024towards}, and Medusa~\cite{medusa}. 

\subsubsection{Fuzzing throughput}
\begin{figure}[]
    \centering
    \includegraphics[width=0.8\linewidth]{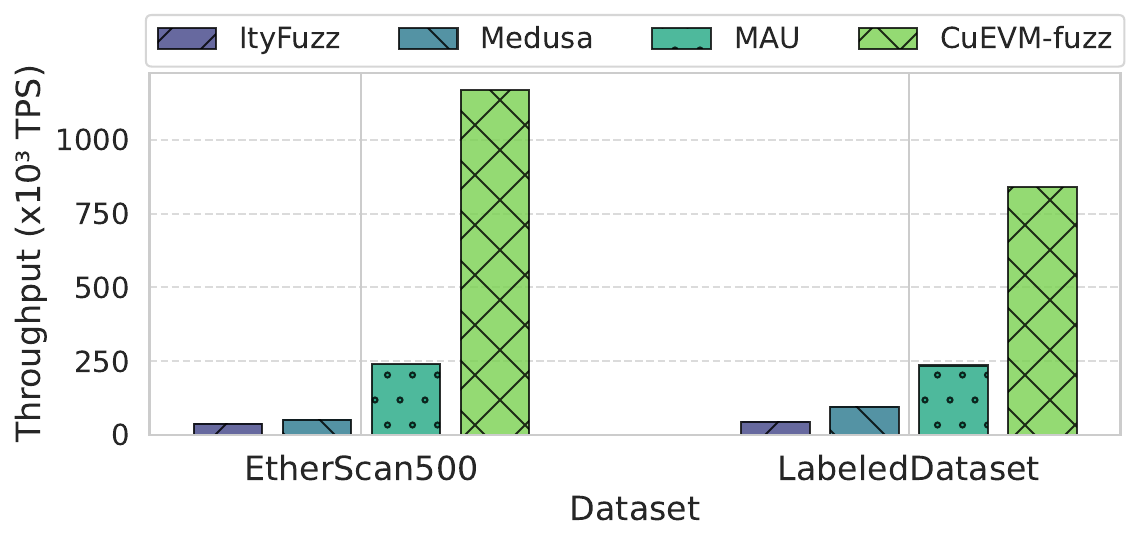}
    \caption{Fuzzing throughput}
    \Description{}
    \label{fig:tx_execution_speed}
\end{figure}

A key performance metric for fuzzing is the seed execution speed: the more seeds a fuzzer runs, the faster it can uncover program paths and bugs. Figure~\ref{fig:tx_execution_speed} compares the end-to-end fuzzing seed execution speed for \textsf{LabeledDataset} and \textsf{Etherscan500} datasets. We exclude Smartian due to its low performance as reported in ~\cite{MAUchen2024towards, ItyfuzzShou2023}. We skip the large-scale datasets \textsf{RecentContract} and \textsf{PopularContract} because not all tools can process them successfully. It can be seen that our fuzzer achieves significantly higher throughput, up to $23\times$ faster than the fastest CPU baseline (Medusa) and
$4.8\times$ faster than the state-of-the-art GPU baseline (MAU).

\begin{figure*}[]
    \centering
        \begin{subfigure}[b]{0.32\linewidth}
            \includegraphics[width=\linewidth]{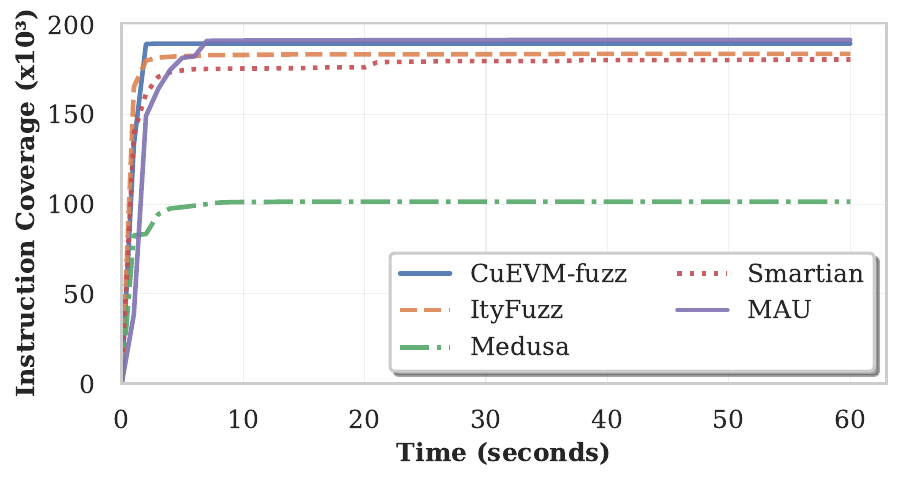}
            \caption{LabeledDataset}
        \end{subfigure}
        \begin{subfigure}[b]{0.32\linewidth}
            \includegraphics[width=\linewidth]{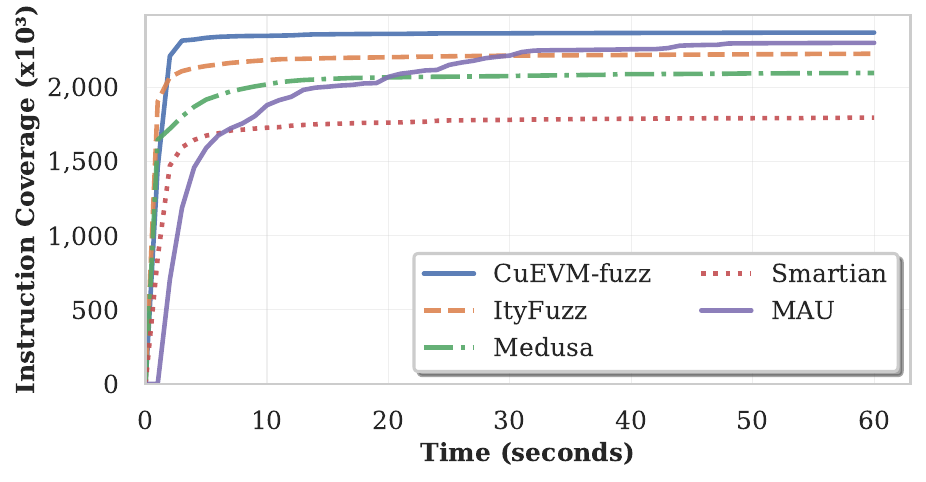}
            \caption{EtherScan500}
        \end{subfigure}
        \begin{subfigure}[b]{0.32\linewidth}
            \includegraphics[width=\linewidth]{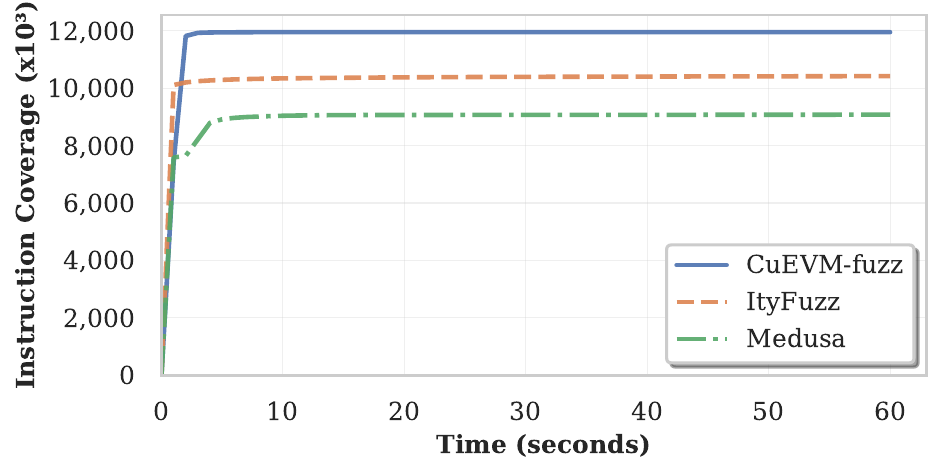}
            \caption{RecentContract}
        \end{subfigure}
        \hfill
        \caption{Instruction coverage on different datasets over time, measured every second
        }
        \Description{}
    \label{fig:instruction_coverage}
\end{figure*}
\subsubsection{Code coverage}
Figure~\ref{fig:instruction_coverage} compares the instruction coverage of the fuzzers over time on the three datasets. On \textsf{RecentContract}, MAU failed to generate the PTX files for 3,335, or 90\% of the contracts, and Smartian can only execute 21 contracts. Thus, we do not include the results of these two baselines. We further exclude contracts that \ityfuzz{} cannot process to ensure a fair comparison. The reported results cover 3,354 contracts supported by all remaining fuzzers. The result shows that \fuzztoolname{} achieves the best coverage on both the  \textsf{EtherScan500} and \textsf{RecentContract} within the first two seconds. Compared to the second-best baseline, MAU, it achieves 4\% more coverage in $30\times$ shorter time. The gap between \fuzztoolname{} and \ityfuzz{} arises from features such as extracting senders from the source code. Medusa underperformed mainly because it does not fuzz fallback functions.

\subsubsection{Bug finding performance}
\label{sec:bug_finding_exp}
\begin{figure}[]
    \begin{subfigure}[b]{0.45\linewidth}
        \centering
        \includegraphics[width=\linewidth]{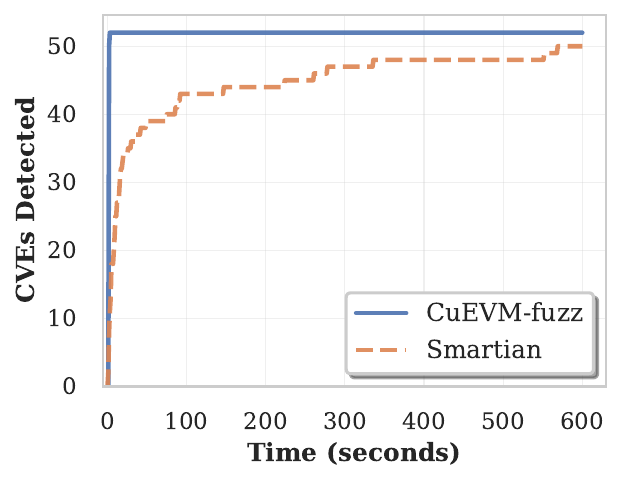}
        \caption{CVEs detected over time on \textsf{LabeledDataset}}
        \label{fig:cve_over_time}
    \end{subfigure}
    \hfill
    \begin{subfigure}[b]{0.45\linewidth}
        \centering
        \includegraphics[width=\linewidth]{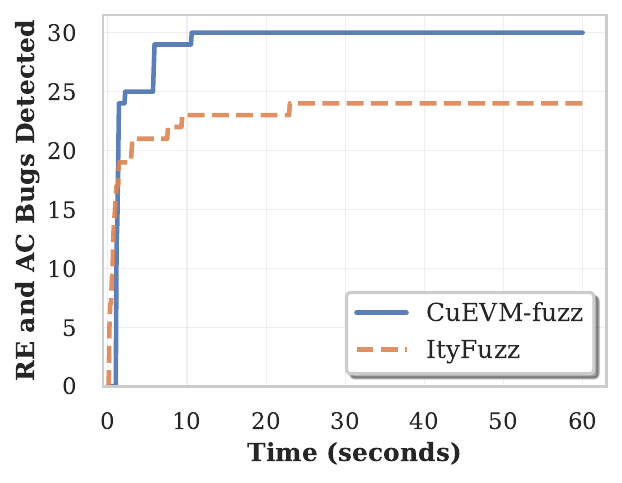}
        \caption{AC and RE bugs detected over time on \textsf{EtherScan500}}
        \label{fig:AC_RE_overtime}
    \end{subfigure}

    \caption{Detection speed comparison with the baselines}
    \Description{}
    \label{fig:bug_finding_exp}
\end{figure}

We compare our bug finding performance against \ityfuzz{} and Smartian. We exclude MAU from this experiment since its released artifact does not have a bug detector, and the authors have indicated via email that it remains proprietary. Medusa is excluded because it only has assertion-based oracles. For unlabeled datasets, we confirm bugs manually by mapping bytecode back to Solidity source, spending up to five minutes per report.

On \textsf{LabeledDataset}, Ityfuzz, \fuzztoolname{}, and Smartian reported the same set of 18 true positive labeled RE bugs. For bug detection speed, \fuzztoolname{} reported 51 CVEs in two seconds, which takes Smartian 10 minutes to achieve, as reflected in Figure~\ref{fig:cve_over_time}.
On unlabeled datasets, we report the bug-finding performance of \textsf{Etherscan500} and \textsf{RecentContract} in Table~\ref{tab:bug_detection_etherscan_recentcontract}, where the \textsf{Bugs} row shows the number of true positives and the \textsf{FP} row indicates the number of false positives. \fuzztoolname{} finds $1.75\times$ more IB bugs in one minute than Smartian does in 10 minutes. It identifies eight exploitable LE bugs caused by constructor name mismatches, $2\times$ more than Smartian. They are critical bugs that allow attackers to gain full control of the contract and withdraw all the funds.
\fuzztoolname{} finds 20 arbitrary call bugs in this well-fuzzed dataset. Interestingly, some of these bugs are from the ERC827 token standard draft that was deemed unsafe and had been removed~\cite{eip827_arbitrary_call}.
For a comparison of bug detection speed between \fuzztoolname{} and \ityfuzz{}, we use \textsf{Etherscan500} and the two types of bugs, Arbitrary Call and Reentrancy (AC and RE) that \ityfuzz{} supports. The speed of both tools is reported in Figure~\ref{fig:AC_RE_overtime}. It shows that \fuzztoolname{} reports more bugs in two seconds compared to \ityfuzz{} performance in a minute. %

\begin{table}[t]
\centering
\caption{Bug detection results on EtherScan500 and RecentContract. IB, LE, AC, RE indicate Integer Overflow/Underflow, Leaking Ether, Arbitrary Call, and Reentrancy.}
\label{tab:bug_detection_etherscan_recentcontract}
\small
\resizebox{\linewidth}{!}{%
\begin{tabular}{llcccc|cccc}
\toprule
& & \multicolumn{4}{c|}{\textsf{EtherScan500}} & \multicolumn{4}{c}{\textsf{RecentContract}} \\
\cmidrule(lr){3-6} \cmidrule(lr){7-10}
& & IB & LE & AC & RE & IB & LE & AC & RE \\
\midrule
\multirow{2}{*}{Smartian} & Bugs & 169 & 4 & - & 0 & - & - & - & - \\
 & FP & 0 & 2 & - & 0 & - & - & - & - \\
\midrule
\multirow{2}{*}{ItyFuzz} & Bugs & - & - & 18 & 6 & - & - & 0 & 0 \\
 & FP & - & - & 0 & 193 & - & - & 0 & 3 \\
\midrule
\multirow{2}{*}{\fuzztoolname{}} & Bugs & 295 & 8 & 20 & 10 & 5 & 6 & 0 & 0 \\
 & FP & 0 & 0 & 0 & 0 & 1 & 16 & 1 & 0 \\
\bottomrule
\end{tabular}
}
\end{table}

\subsubsection{Vulnerability detection in the wild}
To further evaluate the practical security relevance of \fuzztoolname{} on active projects, we leverage its high throughput to perform large-scale scanning on the \textsf{PopularContract} dataset (10,000 contracts each from Ethereum and Binance Smart Chain, BSC). We run \fuzztoolname{} and \ityfuzz{} with a time budget of 10 seconds per contract.

In total, \fuzztoolname{} uncovers several high-risk bugs in 115 contracts (42 from Ethereum and 73 from BSC). Among them, 89 contracts contain IB bugs, 11 contain RE bugs, 9 contain AC bugs, and the remainder contain LE bugs. Four contracts exhibit two bug types (AC and LE). Notably, bugs observed in the \textsf{EtherScan500} still exist in the \textsf{PopularContract}. One prime example is the ERC-827 token, which is still used by projects despite being reported as unsafe~\cite{eip827_arbitrary_call}. Many projects, especially on BSC, continue to use legacy code before Solidity 0.8 and remain vulnerable to IB. The majority of the affected projects were ERC-20 tokens that lack long-term maintainers. Despite using Solidity 0.8, 16 contracts still suffer from IB due to the use of \textit{unchecked} blocks, which disable compiler-injected arithmetic checks to reduce gas costs in frequently executed operations. The true positives above were filtered from 303 contracts reported by \fuzztoolname{}. In comparison, \ityfuzz{} discovers true vulnerabilities in 18 projects (12 AC and 6 RE)%
. Among them, 1 RE and 4 AC bugs are not detected by \fuzztoolname{}. We note that \ityfuzz{}'s bug oracles exhibit a higher false positive rate (256 out of 274 alarms are false positives). %

We adopt best-effort responsible disclosure for the affected contracts, emailing the contacts listed on project websites or cryptocurrency trackers and, where none exist, messaging the owner addresses through Blockscan. %
Three teams acknowledged the issues, having added user warnings or already being aware of them. Some projects with AC bugs carry warnings on Etherscan advising users not to grant token permissions,\footnote{\url{https://archive.ph/beCZt}} five contracts were already labeled ``phishing/hack'' by blockchain explorers, and some projects had migrated to new contracts following the same findings in past professional audits,\footnote{\url{https://paladinsec.co/projects/unsellable-nfts/}} which confirms that \fuzztoolname{} surfaces bugs that professional services also report.

\subsubsection{False positive analysis}
The source of false positives is mainly caused by incomplete information observed by the oracles, similar to existing fuzzers. In IB bugs, the most common case is the intended overflow used by customized math libraries with intended overflow wraparound to implement $2^{256}$ modulo arithmetic. Furthermore, some IB bugs require the contract owners to set certain parameters to be extremely large, which only occurs by accidentally calling the functions outside their intended usage. For RE bugs, the most common false-positive pattern is harmless reentrancy, where reentrancy happens, but it does not result in a trivial exploit leading to direct fund loss. One example is the deposit of user's controlled assets, where reentrancy results in repeating the deposit actions multiple times. %
Ityfuzz has a high false positive rate for RE bugs because its default oracle only detects call loops, not post-call state writes, causing false positives in harmless external or read-only calls.%
\subsubsection{Fuzzing scalability}
\begin{figure}[]
    \centering
    \includegraphics[width=0.98\linewidth]{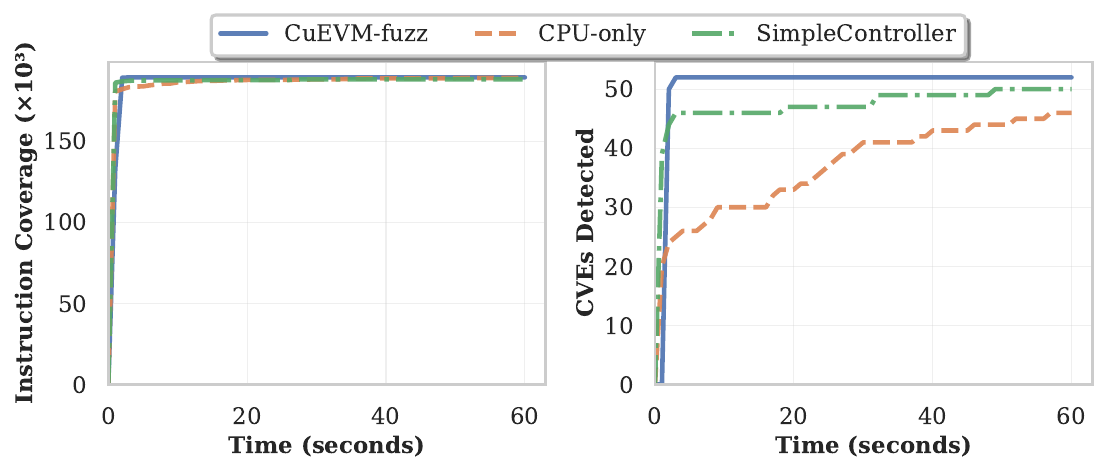}
    \caption{Ablation study
    on \textsf{LabeledDataset} }
    \label{fig:ablation_study_quality}
\end{figure}

\begin{figure}[]
    \centering
    \includegraphics[width=0.96\linewidth]{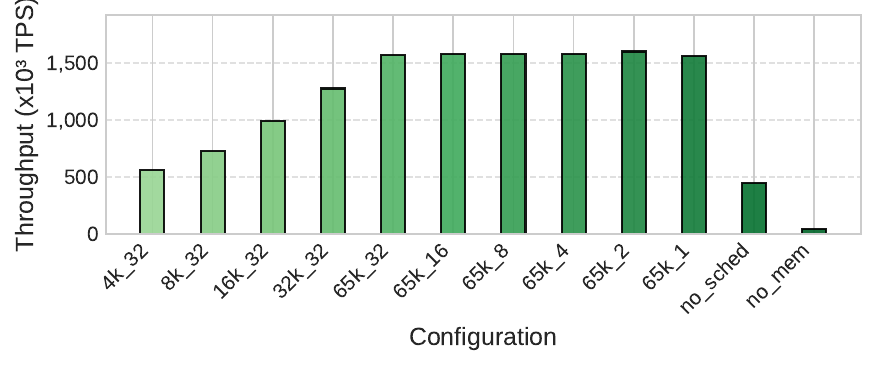}
    \caption{Throughput of different \fuzztoolname{} variants on a subset of \textsf{Etherscan500}}
    \Description{}
    \label{fig:ablation_study_speed}
\end{figure}

We study the scalability of \fuzztoolname{} by varying configurations and components that affect execution speed, such as the number of GPU threads per kernel and workers for the fuzzing controller on CPU. In addition, we introduce two variants to validate the transaction scheduling policy and the choice of data structure design in Section~\ref{sec:coalesced_memory_intro}. For manageable runtime, we sort the \textsf{EtherScan500} dataset by bytecode size and sample 50 representative contracts evenly spaced in the sorted list. The end-to-end fuzzing throughput results are reported in Figure~\ref{fig:ablation_study_speed}. The configuration format is \{GPU\_threads\}\_\{CPU\_threads\}, for example, configuration \textit{8k\_32} means the fuzzer uses 8,192 GPU threads and 32 goroutines on the CPU. We vary the GPU threads from 4k to 65k. We ran the configuration \textit{65k\_1} for the \textit{no\_sched} (remove warp-based scheduling, heighten thread divergence) and \textit{no\_mem} (remove the data preallocation scheme and coalesced design) variants. The results show that \fuzztoolname{} scales well with the number of GPU threads, while CPU workers contribute little to end-to-end performance. This suggests that the bottleneck of end-to-end fuzzing throughput remains on the GPU side, indicating the potential of multi-GPU systems for further speedup. Both \textit{no\_sched} and \textit{no\_mem} perform poorly compared to other \fuzztoolname{} variants, with \textit{no\_mem} showing the worst performance. This indicates that memory allocation and coalesced access patterns are the most important factors for a performant GPU-native execution engine solution, as many instances access memory at high frequency.

\subsubsection{Fuzzing-quality ablation}
We evaluate the fuzzing quality of two variants of \fuzztoolname{}, namely \textsf{CPU-only} and \textsf{SimpleController}. The former simulates the execution of \toolname{} on CPU with goroutines to validate the need for a GPU-native execution engine. The latter uses a basic fuzzing controller without function-dependency extraction from Slither. Figure~\ref{fig:ablation_study_quality} shows the performance of the two variants in terms of code coverage and bug detection. Although similar coverage is achieved quickly by both variants, their bug-detection performance exhibits a different trend. The \textsf{CPU-only} variant lags behind by a large margin compared to the \textsf{SimpleController} version that still uses GPUs for executing transaction. The result strengthens our motivation for introducing GPU acceleration. Since instruction coverage can be saturated early, the main objective of fuzzing is to further mutate inputs on covered paths and run bug oracles. \fuzztoolname{} accelerates both of these steps on GPUs, resulting in faster bug discovery.

\subsection{Incident Response}
We extend \toolname{} to implement a prototype that searches for rescue transactions with the most value (see Section~\ref{sec:time-critical-ronin-use-case} for an example). We introduce the \textit{Extract Value Oracle}, a universal, project-agnostic oracle that measures the ETH balance of a given account after transaction execution and returns the GPU thread index yielding the maximum observed balance. Project-specific oracles can likewise track a particular token balance, which requires knowledge of the contracts and tokens involved. We provide implementation details in our public artifact.
Given contract call data encoding
$N$ parameters, the prototype uses the contract's ABI to identify optimizable parameters (e.g., integers or addresses). It then mutates the call data, similar to \fuzztoolname{} to run parallel tests, applying a two-phase random search~\cite{zabinsky2003stochastic}. The oracle selects the thread with the highest attacker balance post-execution, deriving transaction data that maximizes extracted value. Within one second, the prototype finds a transaction extracting an additional 9,000 USD (3.68 ETH) over the actual optimized exploit transaction, which was submitted 49 minutes after the optimizable exploit.

\section{Related Work}
\label{sec:related_work}

\textbf{Parallel database transaction processing.}
Main-memory databases scale throughput on multicore hardware through optimistic
concurrency control and deterministic execution, such as Calvin~\cite{thomson2012calvin}
and Aria~\cite{lu2020aria}. GPUs have also been used to accelerate data processing,
from relational operators~\cite{he2008relational} and analytical queries~\cite{shanbhag2020study}
to high-throughput transaction execution~\cite{he2011high}, where a recurring lesson
is that data placement and host-device transfer usually bound
performance. %

\textbf{Parallel smart contract execution.}
The sequential execution model of EVM limits the blockchain throughput. Monad~\cite{monad} introduces an optimistic execution model that executes transactions in parallel, and retries a transaction if its input storage slots are modified by other transactions. Block-STM~\cite{gelashvili2023block}, FAFO \cite{zarick2025fafo}, and DMVCC \cite{qi2023smart} propose concurrency control algorithms at the transaction and block level. ParallelEVM \cite{lin2025parallelevm} proposes concurrency control algorithms at the operation level. Spectrum~\cite{chen2024spectrum} and Seer~\cite{zhang2024seer} apply techniques in concurrency control and branch prediction, unlocking parallel execution potential in smart contracts. These works target parallel blockchain execution on CPUs under concurrency control. CuEVM is complementary, targeting independent, non-conflicting workloads via GPU data-parallelism.

\textbf{Smart contract fuzzing.}
Inspired by the success of fuzzing in traditional software testing~\cite{fioraldi2020afl++},  smart contract fuzzing has become one of the standard methods to find bugs and protect assets on blockchains. It has evolved from simple binary fuzzing with single transactions~\cite{nguyen2020sfuzz} to
sequence-aware transaction fuzzing~\cite{qian2024mufuzz} and using
approaches that leverage machine learning~\cite{he2019learning_ILF_fuzzing} and data flow information~\cite{Smartianchoi2021}. Most fuzzers focus on enhancing input mutation while executing transactions on CPU-based EVMs. Recent fuzzers improve execution speed by using snapshots to avoid reexecuting costly transactions~\cite{ItyfuzzShou2023} or by using GPUs~\cite{MAUchen2024towards}.
MAU~\cite{MAUchen2024towards} transpiles EVM bytecodes into PTX assembly to emulate EVM execution on GPUs. However, MAU does not fully support EVM semantics, failing over 80\% of the Ethereum Test Suite, and relies on expensive CPU-side evaluation for fuzzing. %

\section{Conclusion}
\label{sec:conclusion}
We present \toolname{}, a GPU-native and semantically complete EVM execution engine. \toolname{} treats transaction execution as a batched, data-parallel workload and applies data-management techniques such as a columnar state layout to achieve up to 8.3 million ERC20 transfers per second. Our scalability study shows that throughput grows with the number of parallel instances and that the bottleneck remains on the GPU side, suggesting that the current performance can grow higher with powerful GPUs or under multi-GPUs settings.  We demonstrate the practicality and performance of \toolname{} by designing a smart contract fuzzer, which achieves up to 1.1 million end-to-end executions per second and detects high-risk bugs in 115 deployed contracts. We show \toolname{}'s generality by implementing an incident response application which can find high-value rescue transactions within one second.

\bibliographystyle{IEEEtran}
\bibliography{main}

\begin{thebibliography}{10}
\providecommand{\url}[1]{#1}
\csname url@samestyle\endcsname
\providecommand{\newblock}{\relax}
\providecommand{\bibinfo}[2]{#2}
\providecommand{\BIBentrySTDinterwordspacing}{\spaceskip=0pt\relax}
\providecommand{\BIBentryALTinterwordstretchfactor}{4}
\providecommand{\BIBentryALTinterwordspacing}{\spaceskip=\fontdimen2\font plus
\BIBentryALTinterwordstretchfactor\fontdimen3\font minus \fontdimen4\font\relax}
\providecommand{\BIBforeignlanguage}[2]{{%
\expandafter\ifx\csname l@#1\endcsname\relax
\typeout{** WARNING: IEEEtran.bst: No hyphenation pattern has been}%
\typeout{** loaded for the language `#1'. Using the pattern for}%
\typeout{** the default language instead.}%
\else
\language=\csname l@#1\endcsname
\fi
#2}}
\providecommand{\BIBdecl}{\relax}
\BIBdecl

\bibitem{defillama}
``{DefiLlama}: {DeFi TVL aggregator.}'' \url{https://defillama.com/}, 2025, (Accessed May 15, 2026).

\bibitem{wood2014ethereum}
G.~Wood \emph{et~al.}, ``Ethereum: A secure decentralised generalised transaction ledger,'' \emph{Ethereum project yellow paper}, vol. 151, no. 2014, pp. 1--32, 2014.

\bibitem{chen2026mevisor}
W.~Chen and X.~Luo, ``Mevisor: High-throughput mev discovery in dexs with gpu parallelism.'' in \emph{NDSS}, 2026.

\bibitem{meng2026houston}
D.~Meng, F.~Gritti, R.~McLaughlin, N.~Ruaro, I.~Grishchenko, C.~Kruegel, and G.~Vigna, ``Houston: Real-time anomaly detection of attacks against ethereum defi protocols.'' in \emph{NDSS}, 2026.

\bibitem{MAUchen2024towards}
W.~Chen, X.~Luo, H.~Cai, and H.~Wang, ``Towards smart contract fuzzing on gpus,'' in \emph{2024 IEEE Symposium on Security and Privacy (SP)}.\hskip 1em plus 0.5em minus 0.4em\relax IEEE, 2024, pp. 2255--2272.

\bibitem{ItyfuzzShou2023}
C.~Shou, S.~Tan, and K.~Sen, ``Ityfuzz: Snapshot-based fuzzer for smart contract,'' in \emph{Proceedings of the 32nd ACM SIGSOFT International Symposium on Software Testing and Analysis}, 2023, pp. 322--333.

\bibitem{shanbhag2020study}
A.~Shanbhag, S.~Madden, and X.~Yu, ``A study of the fundamental performance characteristics of gpus and cpus for database analytics,'' in \emph{Proceedings of the 2020 ACM SIGMOD international conference on Management of data}, 2020, pp. 1617--1632.

\bibitem{etherscan}
``{Etherscan}: {The Ethereum Blockchain Explorer},'' \url{https://etherscan.io/}, 2025, (Accessed May 15, 2026).

\bibitem{blockchain-immitation-game}
K.~Qin, S.~Chaliasos, L.~Zhou, B.~Livshits, D.~Song, and A.~Gervais, ``The blockchain imitation game,'' in \emph{32nd USENIX Security Symposium (USENIX Security 23)}, 2023, pp. 3961--3978.

\bibitem{mohan2017tutorial}
C.~Mohan, ``Tutorial: blockchains and databases,'' \emph{Proceedings of the VLDB Endowment}, vol.~10, no.~12, pp. 2000--2001, 2017.

\bibitem{dinh2018untangling}
T.~T.~A. Dinh, R.~Liu, M.~Zhang, G.~Chen, B.~C. Ooi, and J.~Wang, ``Untangling blockchain: A data processing view of blockchain systems,'' \emph{IEEE transactions on knowledge and data engineering}, vol.~30, no.~7, pp. 1366--1385, 2018.

\bibitem{ruan2021blockchains}
P.~Ruan, T.~T.~A. Dinh, D.~Loghin, M.~Zhang, G.~Chen, Q.~Lin, and B.~C. Ooi, ``Blockchains vs. distributed databases: Dichotomy and fusion,'' in \emph{Proceedings of the 2021 International Conference on Management of Data}, 2021, pp. 1504--1517.

\bibitem{ge2022hybrid}
Z.~Ge, D.~Loghin, B.~C. Ooi, P.~Ruan, and T.~Wang, ``Hybrid blockchain database systems: design and performance,'' \emph{Proceedings of the VLDB Endowment}, vol.~15, no.~5, pp. 1092--1104, 2022.

\bibitem{dinh2017blockbench}
T.~T.~A. Dinh, J.~Wang, G.~Chen, R.~Liu, B.~C. Ooi, and K.-L. Tan, ``Blockbench: A framework for analyzing private blockchains,'' in \emph{Proceedings of the 2017 ACM international conference on management of data}, 2017, pp. 1085--1100.

\bibitem{zheng2020application}
G.~Zheng, L.~Gao, L.~Huang, and J.~Guan, ``Application binary interface (abi),'' in \emph{Ethereum Smart Contract Development in Solidity}.\hskip 1em plus 0.5em minus 0.4em\relax Springer, 2020, pp. 139--158.

\bibitem{dally2021evolution}
W.~J. Dally, S.~W. Keckler, and D.~B. Kirk, ``Evolution of the graphics processing unit (gpu),'' \emph{IEEE Micro}, vol.~41, no.~6, pp. 42--51, 2021.

\bibitem{cook2012cuda}
S.~Cook, \emph{CUDA programming: a developer's guide to parallel computing with GPUs}.\hskip 1em plus 0.5em minus 0.4em\relax Newnes, 2012.

\bibitem{gelashvili2023block}
R.~Gelashvili, A.~Spiegelman, Z.~Xiang, G.~Danezis, Z.~Li, D.~Malkhi, Y.~Xia, and R.~Zhou, ``Block-stm: Scaling blockchain execution by turning ordering curse to a performance blessing,'' in \emph{Proceedings of the 28th ACM SIGPLAN Annual Symposium on Principles and Practice of Parallel Programming}, 2023, pp. 232--244.

\bibitem{lin2025parallelevm}
H.~Lin, H.~Feng, Y.~Zhou, and L.~Wu, ``Parallelevm: Operation-level concurrent transaction execution for evm-compatible blockchains,'' in \emph{Proceedings of the Twentieth European Conference on Computer Systems}, 2025, pp. 211--225.

\bibitem{tophack}
``{REKT Database}: {Top Crypto Hacks},'' \url{https://de.fi/rekt-database}, 2025, (Accessed May 15, 2026).

\bibitem{AFL}
``{AFL}: {American Fuzzy Lop},'' \url{https://github.com/google/AFL}, 2025, (Accessed May 15, 2026).

\bibitem{gao2025fucci}
X.~Gao, Z.~Liu, Y.~Shen, H.~Li, Y.~Liu, H.~Xiao, Y.~Peng, and J.~Cui, ``Fucci: Database transaction fuzzing via random conflict construction and multilevel constraint solving,'' \emph{Proceedings of the VLDB Endowment}, vol.~18, no.~6, pp. 1879--1891, 2025.

\bibitem{nguyen2020sfuzz}
T.~D. Nguyen, L.~H. Pham, J.~Sun, Y.~Lin, and Q.~T. Minh, ``sfuzz: An efficient adaptive fuzzer for solidity smart contracts,'' in \emph{Proceedings of the ACM/IEEE 42nd international conference on software engineering}, 2020, pp. 778--788.

\bibitem{Smartianchoi2021}
J.~Choi, D.~Kim, S.~Kim, G.~Grieco, A.~Groce, and S.~K. Cha, ``Smartian: Enhancing smart contract fuzzing with static and dynamic data-flow analyses,'' in \emph{2021 36th IEEE/ACM International Conference on Automated Software Engineering (ASE)}.\hskip 1em plus 0.5em minus 0.4em\relax IEEE, 2021, pp. 227--239.

\bibitem{torres2021confuzzius}
C.~F. Torres, A.~K. Iannillo, A.~Gervais, and R.~State, ``Confuzzius: A data dependency-aware hybrid fuzzer for smart contracts,'' in \emph{2021 IEEE European Symposium on Security and Privacy (EuroS\&P)}.\hskip 1em plus 0.5em minus 0.4em\relax IEEE, 2021, pp. 103--119.

\bibitem{your-exploit-is-mine}
Z.~Zhang, Z.~Lin, M.~Morales, X.~Zhang, and K.~Zhang, ``Your exploit is mine: Instantly synthesizing counterattack smart contract,'' in \emph{32nd USENIX Security Symposium (USENIX Security 23)}, 2023, pp. 1757--1774.

\bibitem{whitehatstats}
``Whitehat rescues,'' \url{https://safeharbor.securityalliance.org/whitehats}, 2025, (Accessed May 5, 2026).

\bibitem{he2011high}
B.~He and J.~X. Yu, ``High-throughput transaction executions on graphics processors,'' \emph{Proceedings of the VLDB Endowment}, vol.~4, no.~5, pp. 314--325, 2011.

\bibitem{pearson2019evaluating_gpu_mem}
C.~Pearson, A.~Dakkak, S.~Hashash, C.~Li, I.-H. Chung, J.~Xiong, and W.-M. Hwu, ``Evaluating characteristics of cuda communication primitives on high-bandwidth interconnects,'' in \emph{Proceedings of the 2019 ACM/SPEC International Conference on Performance Engineering}, 2019, pp. 209--218.

\bibitem{abadi2008column}
D.~J. Abadi, S.~R. Madden, and N.~Hachem, ``Column-stores vs. row-stores: how different are they really?'' in \emph{Proceedings of the 2008 ACM SIGMOD international conference on Management of data}, 2008, pp. 967--980.

\bibitem{fauzia2015characterizing_gpu_coalesced}
N.~Fauzia, L.-N. Pouchet, and P.~Sadayappan, ``Characterizing and enhancing global memory data coalescing on gpus,'' in \emph{2015 IEEE/ACM International Symposium on Code Generation and Optimization (CGO)}.\hskip 1em plus 0.5em minus 0.4em\relax IEEE, 2015, pp. 12--22.

\bibitem{gpu_coalesced_memory}
``{CUDA C++ Programming Guide},'' \url{https://docs.nvidia.com/cuda/cuda-c-programming-guide/##device-memory-accesses}, 2025, (Accessed May 15, 2026).

\bibitem{winter2021dynamic_allocate}
M.~Winter, M.~Parger, D.~Mlakar, and M.~Steinberger, ``Are dynamic memory managers on gpus slow? a survey and benchmarks,'' in \emph{Proceedings of the 26th ACM SIGPLAN Symposium on Principles and Practice of Parallel Programming}, 2021, pp. 219--233.

\bibitem{EIP2929}
V.~Buterin and M.~Swende. (2020, Sep.) {EIP-2929: Gas cost increases for state access opcodes}. \url{https://eips.ethereum.org/EIPS/eip-2929}. Ethereum Improvement Proposals, no. 2929.

\bibitem{cgbn}
Nvidia, ``{CGBN}: {CUDA Accelerated Multiple Precision Arithmetic (Big Num) using Cooperative Groups},'' \url{https://github.com/NVlabs/CGBN}, 2025, (Accessed May 15, 2026).

\bibitem{sha}
``{SHA-2}: {SHA-2 Algorithm Implementations},'' \url{https://github.com/amosnier/sha-2}, 2025, (Accessed May 15, 2026).

\bibitem{bigintC}
``{BigInt}: {Big Integer Library in C},'' \url{https://github.com/983/bigint}, 2025, (Accessed May 15, 2026).

\bibitem{pyecc}
``{PyECC}: Python implementation of {ECC} pairing and curve operations,'' \url{https://github.com/ethereum/py_ecc}, 2025, (Accessed May 15, 2026).

\bibitem{go-ethereum}
``{Go-Ethereum}: {Go implementation of the Ethereum protocol},'' \url{https://github.com/ethereum/go-ethereum}, 2025, (Accessed May 15, 2026).

\bibitem{medusa}
``{Medusa}: {Parallelized, coverage-guided, mutational Solidity smart contract fuzzing, powered by Go-Ethereum},'' \url{https://github.com/crytic/medusa}, 2025, (Accessed May 15, 2026).

\bibitem{gpu_atomic_function}
``{CUDA C++ Programming Guide},'' \url{https://docs.nvidia.com/cuda/cuda-c-programming-guide/##atomic-functions }, 2025, (Accessed May 15, 2026).

\bibitem{wang2019branchkeyAFL}
J.~Wang, Y.~Duan, W.~Song, H.~Yin, and C.~Song, ``Be sensitive and collaborative: Analyzing impact of coverage metrics in greybox fuzzing,'' in \emph{22nd International Symposium on Research in Attacks, Intrusions and Defenses (RAID 2019)}, 2019, pp. 1--15.

\bibitem{solidity_security_concern}
``{Solidity Security Considerations},'' \url{https://docs.soliditylang.org/en/latest/security-considerations.html}, 2025, (Accessed May 15, 2026).

\bibitem{ferreira2020smartbugs}
J.~F. Ferreira, P.~Cruz, T.~Durieux, and R.~Abreu, ``Smartbugs: A framework to analyze solidity smart contracts,'' in \emph{Proceedings of the 35th IEEE/ACM international conference on automated software engineering}, 2020, pp. 1349--1352.

\bibitem{feist2019slither}
J.~Feist, G.~Grieco, and A.~Groce, ``Slither: a static analysis framework for smart contracts,'' in \emph{2019 IEEE/ACM 2nd International Workshop on Emerging Trends in Software Engineering for Blockchain (WETSEB)}.\hskip 1em plus 0.5em minus 0.4em\relax IEEE, 2019, pp. 8--15.

\bibitem{eth-tests}
``{Ethereum Execution Tests (EVM)},'' \url{https://github.com/ethereum/tests}, 2025, (Accessed May 15, 2026).

\bibitem{cuevm-artifact}
``Cuevm experimental artifact,'' \url{https://github.com/cuevm-artifact/cuevm-artifact}, 2026.

\bibitem{goevmlab}
``{Go EVMLab}: {Go EVM Laboratory},'' \url{https://github.com/holiman/goevmlab}, 2025, (Accessed May 15, 2026).

\bibitem{reth}
``{Reth: Rust Ethereum},'' \url{https://github.com/paradigmxyz/reth}, 2025, (Accessed May 15, 2026).

\bibitem{pevm}
``{PEVM}: {RISE Parallel EVM},'' \url{https://github.com/risechain/pevm}, 2025, (Accessed May 15, 2026).

\bibitem{eip827_arbitrary_call}
``{EIP827}: {ERC827 Token Standard (ERC20 Extension)},'' \url{https://github.com/ethereum/EIPs/issues/827 }, 2025, (Accessed May 15, 2026).

\bibitem{zabinsky2003stochastic}
Z.~B. Zabinsky, \emph{Stochastic adaptive search for global optimization}.\hskip 1em plus 0.5em minus 0.4em\relax Springer Science \& Business Media, 2003, vol.~72.

\bibitem{thomson2012calvin}
A.~Thomson, T.~Diamond, S.-C. Weng, K.~Ren, P.~Shao, and D.~J. Abadi, ``Calvin: fast distributed transactions for partitioned database systems,'' in \emph{Proceedings of the 2012 ACM SIGMOD international conference on management of data}, 2012, pp. 1--12.

\bibitem{lu2020aria}
Y.~Lu, X.~Yu, L.~Cao, and S.~Madden, ``Aria: a fast and practical deterministic oltp database,'' \emph{Proceedings of the VLDB Endowment}, vol.~13, no.~12, pp. 2047--2060, 2020.

\bibitem{he2008relational}
B.~He, K.~Yang, R.~Fang, M.~Lu, N.~Govindaraju, Q.~Luo, and P.~Sander, ``Relational joins on graphics processors,'' in \emph{Proceedings of the 2008 ACM SIGMOD international conference on Management of data}, 2008, pp. 511--524.

\bibitem{monad}
``{Monad}: {High-performance Ethereum-compatible L1},'' \url{https://docs.monad.xyz/}, 2025, (Accessed May 15, 2026).

\bibitem{zarick2025fafo}
R.~Zarick, I.~Zhang, D.~Wong, T.~Kim, B.~Pellegrino, M.~Li, and K.~Wong, ``Fafo: Over 1 million tps on a single node running evm while still merkleizing every block,'' \emph{arXiv preprint arXiv:2507.10757}, 2025.

\bibitem{qi2023smart}
X.~Qi, J.~Jiao, and Y.~Li, ``Smart contract parallel execution with fine-grained state accesses,'' in \emph{2023 IEEE 43rd International Conference on Distributed Computing Systems (ICDCS)}.\hskip 1em plus 0.5em minus 0.4em\relax IEEE, 2023, pp. 841--852.

\bibitem{chen2024spectrum}
Z.~Chen, T.~Yang, Y.~Zheng, Z.~Zhang, C.~Jin, and A.~Zhou, ``Spectrum: Speedy and strictly-deterministic smart contract transactions for blockchain ledgers,'' \emph{Proceedings of the VLDB Endowment}, vol.~17, no.~10, pp. 2541--2554, 2024.

\bibitem{zhang2024seer}
S.~Zhang, R.~Cheng, X.~Liu, J.~Xiao, H.~Jin, and B.~Li, ``Seer: Accelerating blockchain transaction execution by fine-grained branch prediction,'' \emph{Proceedings of the VLDB Endowment}, vol.~18, no.~3, pp. 822--835, 2024.

\bibitem{fioraldi2020afl++}
A.~Fioraldi, D.~Maier, H.~Ei{\ss}feldt, and M.~Heuse, ``$\{$AFL++$\}$: Combining incremental steps of fuzzing research,'' in \emph{14th USENIX workshop on offensive technologies (WOOT 20)}, 2020.

\bibitem{qian2024mufuzz}
P.~Qian, H.~Wu, Z.~Du, T.~Vural, D.~Rong, Z.~Cao, L.~Zhang, Y.~Wang, J.~Chen, and Q.~He, ``Mufuzz: Sequence-aware mutation and seed mask guidance for blockchain smart contract fuzzing,'' in \emph{2024 IEEE 40th International Conference on Data Engineering (ICDE)}.\hskip 1em plus 0.5em minus 0.4em\relax IEEE, 2024, pp. 1972--1985.

\bibitem{he2019learning_ILF_fuzzing}
J.~He, M.~Balunovi{\'c}, N.~Ambroladze, P.~Tsankov, and M.~Vechev, ``Learning to fuzz from symbolic execution with application to smart contracts,'' in \emph{Proceedings of the 2019 ACM SIGSAC conference on computer and communications security}, 2019, pp. 531--548.

\end{thebibliography}
\end{document}